\documentclass[fleqn,usenatbib]{mnras}

\usepackage[T1]{fontenc}

\DeclareRobustCommand{\VAN}[3]{#2}
\let\VANthebibliography\thebibliography
\def\thebibliography{\DeclareRobustCommand{\VAN}[3]{##3}\VANthebibliography}

\usepackage{graphicx}	% Including figure files
\usepackage{amsmath}	% Advanced maths commands
\usepackage{amssymb}	% Extra maths symbols
\usepackage{multirow}   % For dealing with table formatting
\usepackage{newtxtext,newtxmath}
\usepackage{gensymb}
\usepackage{hyperref}

\graphicspath{{figs/}}

\setcitestyle{notesep={}}

\title[New $UV$ LF Results at $z\sim8$-15 using {\it JWST} Data over the XDF]{Evolution of the $UV$ LF from $z\sim15$ to $z\sim8$ Using New {\it JWST} NIRCam Medium-Band Observations over the HUDF/XDF}

\author[Bouwens et al.]{Rychard J. Bouwens$^{1}$\thanks{email: bouwens@strw.leidenuniv.nl},
Mauro Stefanon$^{2,3}$, 
Gabriel Brammer$^{4}$,
Pascal A. Oesch$^{4,5}$,
Thomas Herard-Demanche$^{1}$, 
\newauthor
Garth D. Illingworth$^{6}$,
Jorryt Matthee,$^{7}$
Rohan P. Naidu$^{8,9}$,
Pieter G. van Dokkum$^{10}$,
Ivana F. van Leeuwen$^{1}$\\  
$^{1}$Leiden Observatory, Leiden University, NL-2300 RA Leiden, Netherlands\\
$^{2}$Departament d’Astronomia i Astrof{\' i}sica, Universitat de Val{\`e}ncia, C. Dr. Moliner 50, E-46100 Burjassot, Val{\` e}ncia, Spain\\
$^{3}$Unidad Asociada CSIC ”Grupo de Astrof{\' i}sica Extragal{\' a}ctica y Cosmolog{\' i}a” (Instituto de F{ \' i}sica de Cantabria - Universitat de Val{\` e}ncia)\\
$^{4}$Cosmic Dawn Center (DAWN), Niels Bohr Institute, University of Copenhagen, Jagtvej 128, K\o benhavn N, DK-2200, Denmark\\
$^{5}$Department of Astronomy, University of Geneva, Chemin Pegasi 51, 1290 Versoix, Switzerland\\
$^{6}$UCO/Lick Observatory, University of California, Santa Cruz, CA 95064\\
$^{7}$Department of Physics, ETH Z\"{u}urich, Wolfgang-Pauli-Strasse 27, 8093 Z\"{u}urich, Switzerland\\
$^{8}$Center for Astrophysics, Harvard \& Smithsonian, 60 Garden Street, Cambridge, MA 02138, USA\\
$^{9}$MIT Kavli Institute for Astrophysics and Space Research, 77 Massachusetts Ave., Cambridge, MA 02139, USA\\
$^{10}$Astronomy Department, Yale University, 52 Hillhouse Ave, New Haven, CT 06511, USA}

\date{Accepted XXX. Received YYY; in original form ZZZ}

\pubyear{2022}

\begin{document}
\label{firstpage}
\pagerange{\pageref{firstpage}--\pageref{lastpage}}
\maketitle
\begin{abstract}
We present the first constraints on the prevalence of $z>10$ galaxies
in the {\it Hubble} Ultra Deep Field (HUDF) leveraging new NIRCam
observations from JEMS ({\it JWST} Extragalactic Medium-band Survey).
These NIRCam observations probe redward of 1.6$\mu$m, beyond the
wavelength limit of {\it HST}, allowing us to search for galaxies to
$z>10$.  These observations indicate that the highest redshift
candidate identified in the HUDF09 data with {\it HST}, UDFj-39546284,
has a redshift of $z>11.5$, as had been suggested in analyses of the
HUDF12/XDF data. This has now been confirmed with {\it JWST} NIRSpec.
This source is thus the most distant galaxy discovered by {\it HST} in
its $>$30 years of operation.  Additionally, we identify nine other
$z\sim8$-13 candidate galaxies over the HUDF, two of which are new
discoveries that appear to lie at $z\sim11$-12.  We use these results
to characterize the evolution of the $UV$ luminosity function (LF)
from $z\sim15$ to $z\sim8.7$.  While our LF results at $z\sim8.7$ and
$z\sim10.5$ are consistent with previous findings over the HUDF, our
new LF estimates at $z\sim12.6$ are higher than other results in the
literature, potentially pointing to a milder evolution in the $UV$
luminosity density from $z\sim12.6$.  We emphasize that our LF results
are uncertain given the small number of $z\sim12.6$ sources and
limited volume probed.  The new NIRCam data also indicate that the
faint $z\sim8$-13 galaxies in the HUDF/XDF show blue $UV$-continuum
slopes $\beta$ $\sim-2.7$, high specific star formation rates
$\sim$24.5 Gyr$^{-1}$, and high EW ($\sim$1300\AA) [OIII]+H$\beta$
emission, with two $z\sim8.5$ sources showing [OIII]+H$\beta$ EWs of
$\sim$2300\AA.
\end{abstract}

\begin{keywords}
galaxies: evolution -- galaxies: high-redshift -- dark ages, reionization, first stars
\end{keywords}
%%%%%%%%%%%%%%%%%%%%%%%%%%%%%%%%%%%%%%%%%%%%%%%%%%

%%%%%%%%%%%%%%%%% BODY OF PAPER %%%%%%%%%%%%%%%%%%

\section{Introduction} \label{sec:intro}

One of the most interesting frontiers in extragalactic astronomy
concerns the formation and growth of galaxies in the early universe.
Significant open questions exist regarding the efficiency of star
formation in the early universe \citep[e.g.,][]{Behroozi2013,Stefanon2017_optLF}, when
cosmic reionization began in earnest \citep[e.g.,][]{Robertson2015,Bouwens2015_Reion,Robertson2022},
and the overall timeline of stellar assembly
\citep[e.g.][]{Madau2014}.  Leveraging the capabilities of the {\it Hubble Space Telescope (HST)} and {\it Spitzer}, significant progress was made in identifying
galaxies to $z\sim 11$ \citep{Coe2013,Oesch2016,Jiang2021} and probing
the build-up of stellar mass from $z\sim10$
\citep[e.g.,][]{Stark2009,Gonzalez2011,Duncan2014,Grazian2015,Song2016,Bhatawdekar2019,Kikuchihara2020,Furtak2021,Stefanon2021_SMF,Stefanon2022_IRACz10}.  Now, with the initial studies with the {\it James Webb Space Telescope (JWST)}, large numbers of
galaxies have been identified with redshifts in excess of $z\sim10$
\citep[e.g.][]{Adams2022,Naidu2022_z12,Castellano2022_GLASS,Atek2022,Finkelstein2022_z12,Labbe2022,Bradley2022,Hsiao2022,Rodighiero2022} with
plausible redshifts as high as $z\sim17$
\citep[e.g.][]{Donnan2022,Atek2022,Harikane2022_z9to17,Naidu2022_z17,Zavala2022,Yan2022}.

\begin{figure*}
\centering
\includegraphics[width=2\columnwidth]{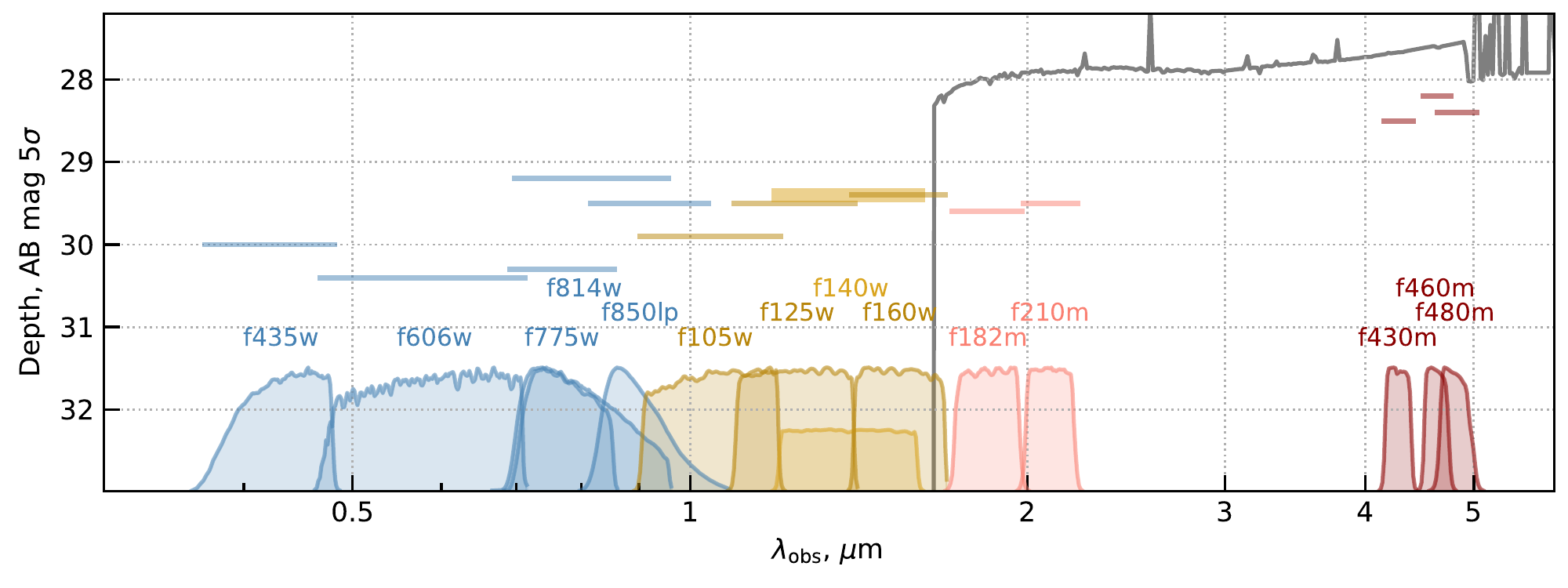}
\caption{Sensitivity ($5\sigma$, AB mag) of the new {\it JWST} medium band observations in F182M, F210M, F430M, F460M, and F480M (\textit{red horizontal lines}) over the HUDF from the JEMS program \citep{Williams2023_JEMS} relative to that available in various filters from the HUDF/XDF optical (\textit{blue lines}) and near-IR (\textit{light brown lines}) observations.  Also shown are the wavelength sensitivity curves for the {\it JWST} medium bands as well as the archival HST observations.  For context, an SED of the $z=8.5$ galaxy over SMACS0723, as fit with \textsc{Bagpipes} \citep{Carnall2018,Carnall2022} and shifted to $z=12$, is also shown (\textit{grey lines}).  Not only are the new {\it JWST} medium band data well matched to the depth of the archival {\it HST} HUDF/XDF observations \citep{Illingworth2013}, but they extend the coverage to redder wavelengths, facilitating both an identification of star-forming galaxies over the HUDF to higher redshifts, i.e., $z>10$, and a characterization of the stellar population and emission line properties of very high-redshift galaxies in the same field.\label{fig:filters}}
\end{figure*}

In the midst of all the excitement surrounding the discovery of many
plausible $z\geq 10$ galaxies, significant questions have persisted
regarding how robust various high redshift candidates from {\it JWST} are.
Of concern has been the limited depth of the imaging data just
blueward of various putative Lyman breaks, the only modest overlap
between the specific sources that make up various $z\geq 8$
selections,\footnote{https://twitter.com/stewilkins/status/1554909423759409153}
and uncertainties in the calibration of the NIRCam zeropoints \citep[e.g.][]{Adams2022}.

One way to substantially improve the robustness of current $z\geq 8$
selections is to perform these selections over those extragalactic fields with
some of the deepest available imaging observations with {\it HST}, and there is clearly no field with deeper archival
observations than the {\it Hubble} Ultra Deep Field (HUDF).  The HUDF has
been the target of more than 1000 hours of observations with {\it HST}
\citep{Thompson2005,Beckwith2006,Bouwens2011_LF,Ellis2013,Koekemoer2013,Illingworth2013,
  Teplitz2013}, probing to $5\sigma$ limiting magnitudes $\geq$30 mag in the bluest three optical bands, to $\sim$30 mag in the F105W band, and to $\sim$29.5 mag otherwise (F850LP, F125W, F140W, and F160W).  This is approximately 2 mag deeper at optical wavelengths than has been available over other fields with early JWST observations, e.g., the Cosmic Evolution Early Release Science (CEERS) program \citep{Finkelstein2017_CEERS}, SMACS0723 \citep{Pontop2022}, or the GLASS parallel field \citep{Treu2022_GLASS}.
  
Here we take advantage of sensitive NIRCam imaging data which have just been
obtained over the {\it Hubble} Ultra Deep Field, as part of the sensitive JEMS
({\it JWST} Extragalactic Medium-band Survey) program \citep{Williams2023_JEMS}.  Imaging
observations from the program were obtained just redward of the
sensitive Wide Field Camera 3 (WFC3)/IR observations from the HUDF09
and HUDF12 programs \citep{Bouwens2011_LF,Ellis2013} and enable searches for star-forming galaxies over
the {\it Hubble} Ultra Deep Field to even higher redshift than can be
readily probed with the WFC3/IR data sets, i.e., to $z>10$ and out to
$z\sim15$.  The observations also allow for a reevaluation of various
candidate $z\geq 10$ galaxies that have already been reported over
that field and a rederivation of $UV$ LF results at $z\geq 8$.  Thanks
to the depth of the available supporting observations over the {\it
  Hubble} Ultra Deep Field (being up to $\sim$2 mag deeper than available over other fields), we would expect these new $UV$ LF results
to be more reliable than most earlier determinations with {\it JWST}.

In \S2, we describe the combined {\it JWST} NIRCam plus {\it HST} imaging data
sets available to search for star-forming galaxies over the {\it Hubble}
Ultra Deep Field as well as our procedures for constructing
source catalogs and performing photometry.  In \S3, we describe our procedures for
identifying star-forming galaxies over the {\it Hubble} Ultra Deep Field,
and then present the samples of sources we derive and compare it to
earlier selections.  In \S4, we use our new samples of $z\sim8$-13
galaxies to quantify the $UV$ LF at $z\geq 8$, characterize the
evolution with cosmic time, and analyze the stellar populations and line emission from fainter star-forming galaxies at $z\geq 8$.  We refer to the {\it HST} F435W, F606W, F775W,
F814W, F850LP, F105W, F125W, F140W, and F160W bands as $B_{435}$,
$V_{606}$, $i_{775}$, $I_{814}$, $z_{850}$, $Y_{105}$, $J_{125}$,
$JH_{140}$, and $H_{160}$, respectively, for simplicity, and to the
{\it JWST} F182M and F210M bands as $HK_{182}$ and $K_{210}$, respectively.
For convenience, we quote results in terms of the approximate
characteristic luminosity $L_{z=3}^{*}$ derived at $z\sim3$ by
\citet{Steidel1999}, \citet{Reddy2009}, and many other studies.  For ease of comparison to other recent extragalactic work, we assume a concordance cosmology with $\Omega_{m}=0.3$, $\Omega_{\Lambda}=0.7$, and $H_0 = 70 \,\textrm{km/s/Mpc}$ throughout.  All SFR and stellar mass results are quoted assuming a \citet{Chabrier2003} initial mass function (IMF).  All
magnitude measurements are given using the AB magnitude system
\citep{OkeGunn1983} unless otherwise specified.

\begin{table}
\centering
\caption{Estimated $5\sigma$ depth [in mag] of the {\it JWST}+{\it HST} data set we utilize over the HUDF/XDF in 0.35$''$ diameter apertures.}
\label{tab:dataset}
\begin{tabular}{c|c|c|c} \hline
Band & $5\sigma$ Depth (mag) \\\hline\hline
{\it HST}/F435W & 30.0 \\
{\it HST}/F606W & 30.4 \\
{\it HST}/F775W & 30.3 \\
{\it HST}/F814W & 29.2 \\
{\it HST}/F850LP & 29.5 \\
{\it HST}/F105W & 29.9 \\
{\it HST}/F125W & 29.5 \\
{\it HST}/F140W & 29.4 \\
{\it HST}/F160W & 29.4 \\ 
{\it {\it JWST}}/F182M & 29.6 \\
{\it {\it JWST}}/F210M & 29.5 \\
{\it {\it JWST}}/F430M & 28.5 \\
{\it {\it JWST}}/F460M & 28.2 \\
{\it {\it JWST}}/F480M & 28.4 \\
Area [arcmin$^2$] & 4.6 \\\hline\hline
\end{tabular}
\end{table}

\section{Data Sets and Photometry}

\subsection{Data Set\label{sec:data}}

Here we make use of the new {\it JWST} NIRCam
observations taken over the {\it Hubble} Ultra Deep Field as part of the JEMS ({\it JWST} extragalactic mediumband survey) medium-band program \citep{Williams2023_JEMS}.  
Obtained as part
of the program were observations in F182M and F210M short wavelength
NIRCam channel and F430M, F460M, and F480M observations taken with the longer wavelength NIRCam channel.  The F182M, F210M, and F480M
observations had total integration times of 7.8 hours each, while the
total integration times for the F430M and F460M observations were 3.9
hours each in duration.  The observations were reduced using
\textsc{grizli} reduction pipeline adapted to handle {\it JWST} NIRCam data (G. Brammer et al.\ 2022, in prep).  \textsc{grizli} includes procedures for
masking the ``snowball'' artefacts and minimizing the impact of $1/f$
noise.  Image combination is treated using \textsc{astrodrizzle} after
converting the WCS information in the image headers to use
\textsc{SIP} format.

The WFC3/IR and Advanced Camera for Surveys (ACS) images used for our
analysis are the XDF images that were generated in 2013 based on all
the overlapping {\it HST} optical and near-IR images that were available at that time and include $\sim$1000 orbits of data
\citep{Illingworth2013}.  These images include deep observations in
the F435W, F606W, F775W, F814W, and F850LP bands with ACS and F105W,
F125W, F140W, and F160W bands with WFC3/IR.

Figure~\ref{fig:filters} and Table~\ref{tab:dataset} summarize the $5\sigma$ depth of the imaging
data sets we utilize for our analysis, and what is striking is how
similar the depths of the available near-IR data sets with {\it HST} are to
the depths of the imaging data sets available with {\it JWST} in the short
wavelength channel, both reaching to $\sim$29.5 mag at $5\sigma$ while
the depth of the medium-band observations in the longer wavelength
channel is 28.2-28.5, sensitive enough to look for prominent line
emission from H$\alpha$ and [OIII]+H$\beta$ in galaxies at $z\sim
5.4$-6.6 and $z\sim7.6$-9.3, respectively \citep{Williams2023_JEMS}.

\begin{figure}
\centering
\includegraphics[width=\columnwidth]{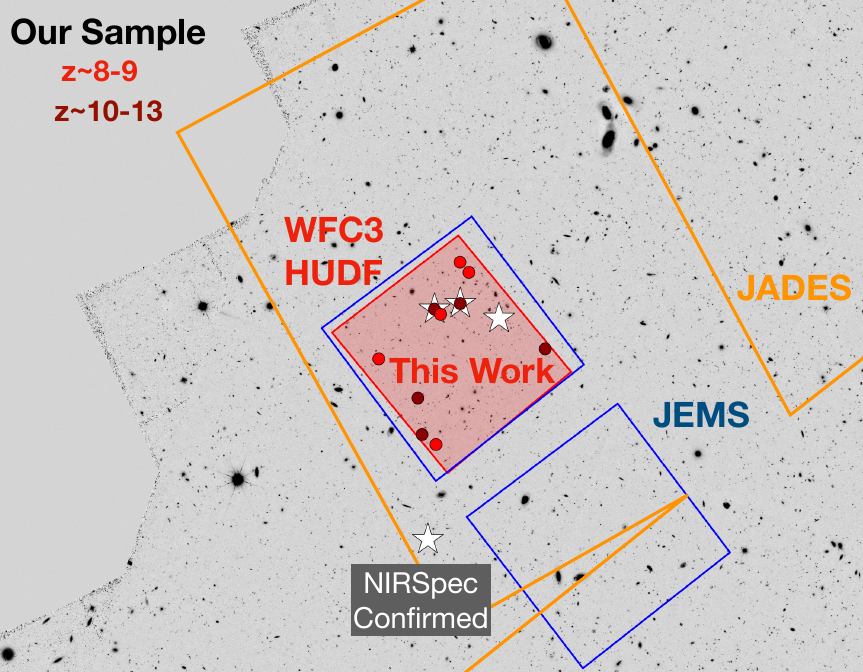}
\caption{Illustration of our search field (\textit{red shaded region}) for $z\sim8$-15 galaxies leveraging the deep JEMS medium-band data (\textit{blue rectangles}) and sensitive WFC3/IR data over the HUDF/XDF from HUDF09+HUDF12 programs (\textit{red rectangle}).  The background image shown in greyscale corresponds to the GOODS South {\it HST} ACS F606W observations \citep{Giavalisco2004_GOODS}.  Also shown are the areas covered by the JADES NIRCam fields (\textit{bracketed by the orange lines}).  The coordinates of $z\sim8$-9 and $z\sim10$-13 candidate galaxies in
our selection are indicated by the red and dark red circles, respectively.  The four white stars indicate the coordinates for the four $z\sim10$-14 galaxies which have been spectroscopically confirmed by the JADES NIRSpec observations in \citet{CurtisLake2022_JADESz13} and \citet{Robertson2022_JADESz13}.\label{fig:layout}}

\end{figure}

We restrict our search to the $\sim$4.6 arcmin$^2$ region on the sky that contains both the deepest WFC3/IR observations from the HUDF09+HUDF12 programs and {\it JWST}/NIRCam medium-band observations from the JEMS program \citep{Williams2023_JEMS}.  The footprint of our 4.6 arcmin$^2$ search field relative to HUDF09+HUDF12 dataset, JEMS field, and JADES NIRCam footprint in the GOODS South is shown in Figure~\ref{fig:layout}.  Also indicated are four spectroscopically-confirmed sources from \citet{CurtisLake2022_JADESz13} and \citet{Robertson2022_JADESz13}.

\begin{figure*}
\centering
\includegraphics[width=1.8\columnwidth]{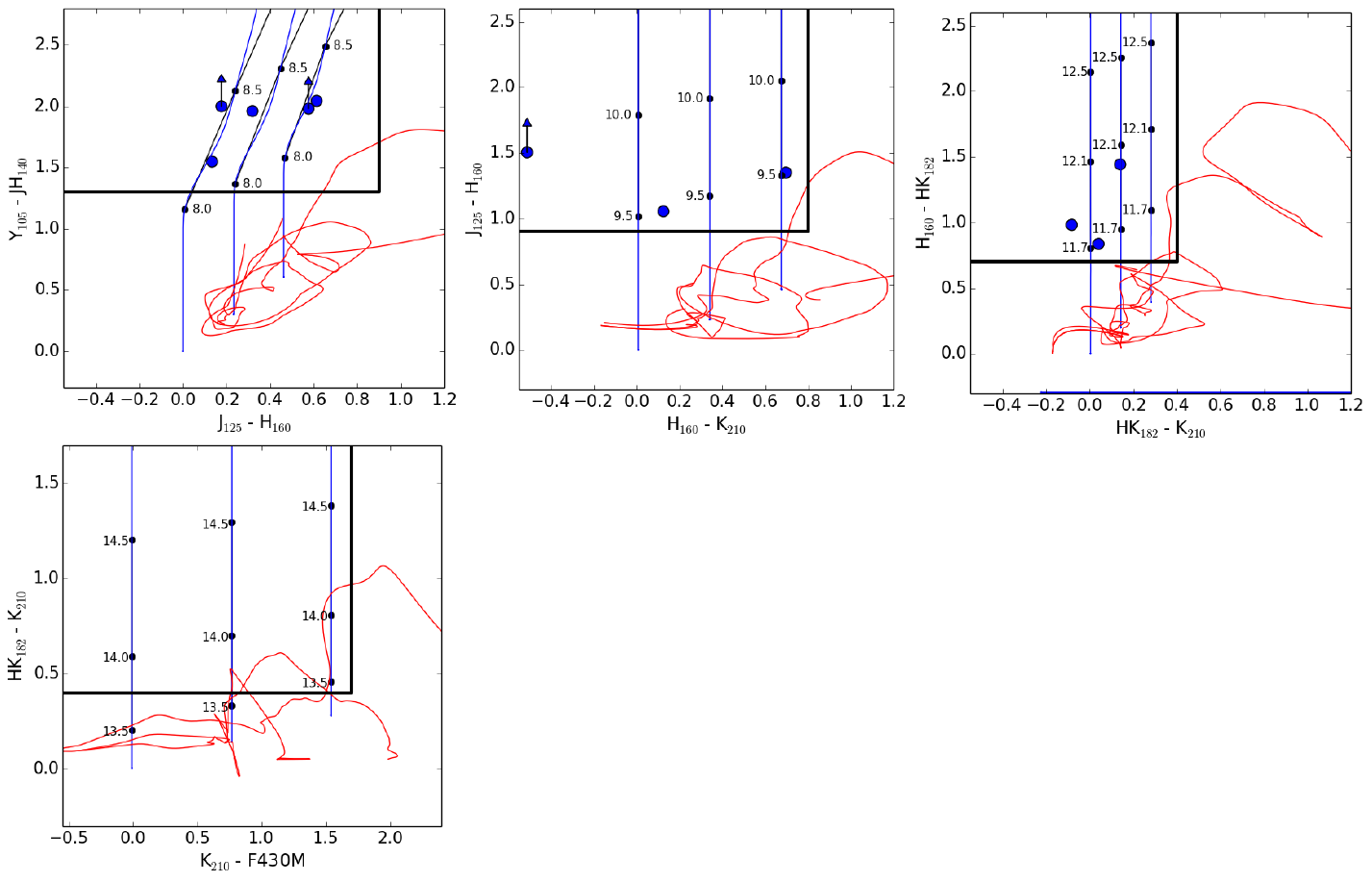}
\caption{Illustration of the main color criteria we use to identify galaxies at $z\sim8$-9 (upper left), $z\sim10$-11 (upper center), $z\sim12$-13 (upper right), and $z\sim14$-15 (lower left).   The thick black lines indicate the boundaries of our
 primary color-color criteria (but note that for our $z\sim8$-9 selection we also include sources with $J_{125}-H_{160}$ colors redder than 0.3 mag and $Y_{105}-JH_{140}$ colors redder than 0.8).  The blue lines indicate the expected colors
  of star-forming galaxies with 100 Myr constant star-formation histories and
  $E(B-V)$ dust extinction of 0, 0.15, and 0.3, with colors at
  specific redshifts indicated by the black dots.  The red lines show
  the expected colors of lower-redshift template galaxies from
  \citet{Coleman1980} out to $z\sim7$.  The solid blue circles show
  the colors of specific sources in our selection.  In cases
  where sources are not detected in a band, they are shown with an
  arrow at the $1\sigma$ limit. \label{fig:lbgcrit}}
\end{figure*}

\begin{figure}
\centering
\includegraphics[width=\columnwidth]{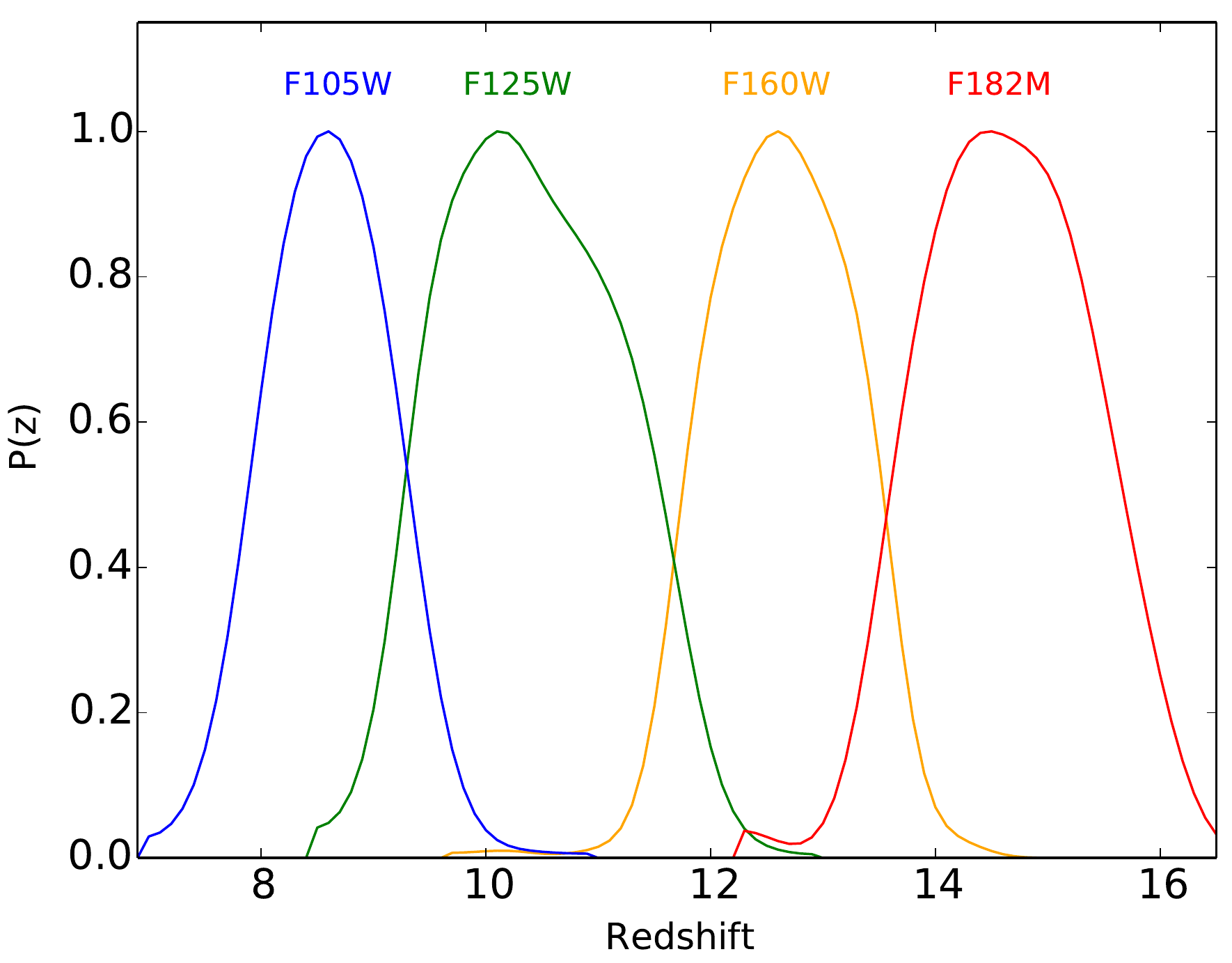}
\caption{Redshift selection functions for our $z\sim8$-9 $Y_{105}$-dropout, $z\sim10$-11 $J_{125}$-dropout,  $z\sim12$-13 $H_{160}$-dropout, and $z\sim14$-15 $HK_{182}$-dropout selections.  The criteria for these selections is described in \S\ref{sec:lbgselect}. The expected redshift distribution of these selections
  is estimated using our selection volume simulations in
  \S\ref{sec:uvlfs} and has a mean value of 8.7, 10.5, 12.6, and 14.7, respectively.}
    \label{fig:zsel}
\end{figure}

\begin{figure*}
\centering
\includegraphics[width=2\columnwidth]{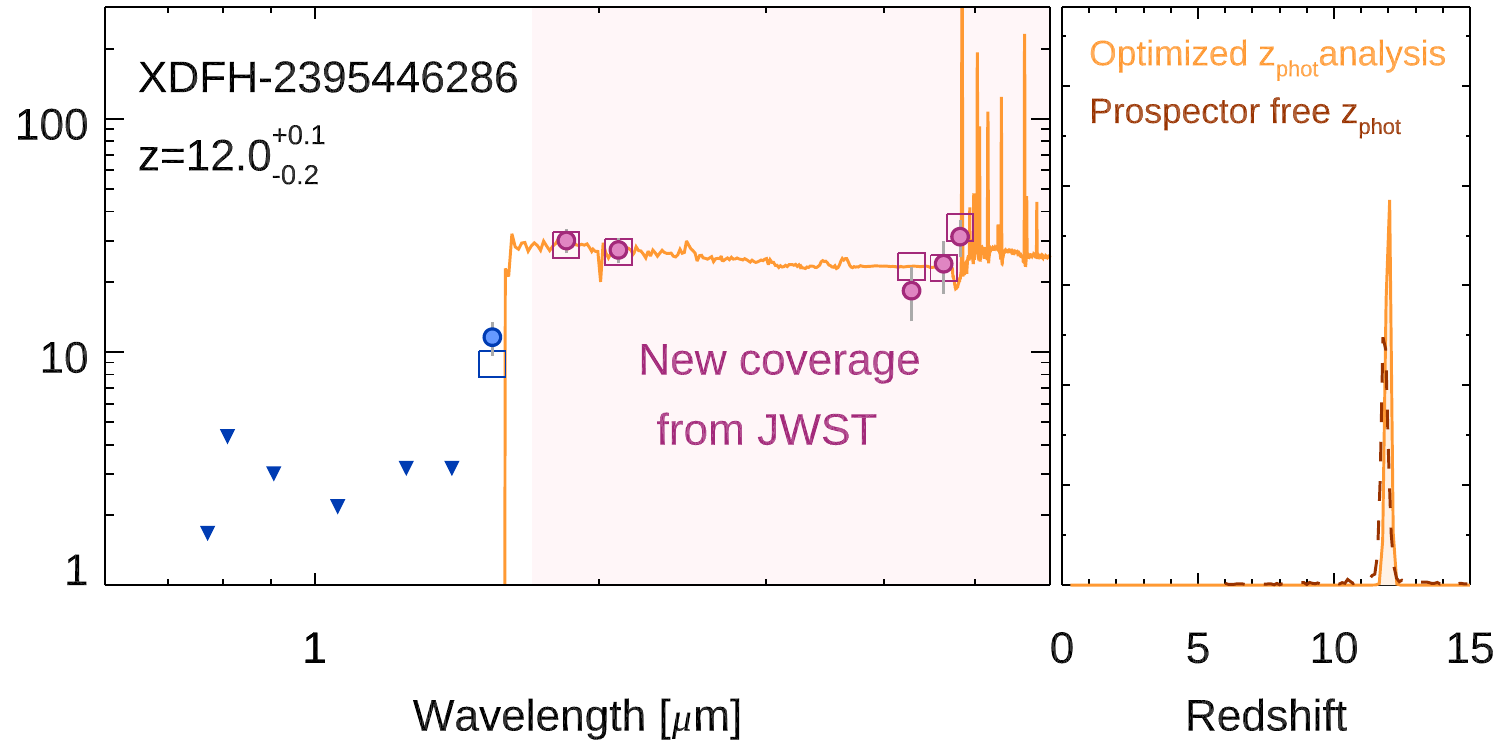}
\caption{(\textit{left}) Fits of a model spectral energy distribution
  to the observed {\it HST} + {\it JWST} photometry available for the
  highest redshift candidate UDFj-39546284 identified over the HUDF
  with {\it HST} \citep{Bouwens2011_Nature}.  Flux measurements
  blueward of 1.6$\mu$m are with {\it HST} (\textit{blue filled points and downward triangles}) and those redward of this
  limit are with {\it JWST} (\textit{magenta filled points}).  The blue downward triangles correspond to 2$\sigma$ upper limits on the fluxes.  The open squares indicate the expected fluxes from the best-fitting SED model.  
  (\textit{right}) Redshift likelihood
  distribution derived for UDFj-39546284 from the \text{EAzY}
  photometric redshift code \citep{Brammer2008} on our photometric
  measurements, and UDFj-39546284 seems to almost certainly have a
  redshift of $z=12.0_{-0.2}^{+0.1}$, as the JADES team has confirmed with spectroscopy \citep{CurtisLake2022_JADESz13} and similar to what \citet{Ellis2013},
  \citet{McLure2013}, \citet{Oesch2013}, and \citet{Bouwens2013}
  inferred using the available {\it HST}+{\it Spitzer} data in 2013.  As
  such, UDFj-39546284 appears to be the most distant galaxy discovered by {\it HST} in its more than 30 years of operation.   Figure~\ref{fig:stamph} shows postage stamp images of this source. \label{fig:sedfit}}
\end{figure*}

\subsection{Source Detection and Photometry}

For source detection and photometry over the HUDF, we make use of the SExtractor code \citep{Bertin1996}.  Source detection is performed
leveraging a square root of $\chi^2$ image \citep{Szalay1999},
constructed from the observations in the F140W, F160W, F182W, and
F210M bands for our $z\sim8$-9 and $z\sim10$-11 selections, observations in the F182W and F210M bands for our $z\sim12$-13 selection, and F210M band for our $z\sim14$-15 selection.  The various images contributing to the detection image are first PSF-matched to the lowest resolution bandpass F160W prior to coaddition.
Color measurements are made inside small scalable \cite{Kron1980} apertures,
using a Kron factor of 1.2.  Flux measurements made in the small
scalable apertures are corrected to total by accounting for the
additional flux in the square root of $\chi^2$ in larger scalable Kron
apertures (Kron factor 2.5) relative to the smaller scalable
apertures.  An additional correction is made to account for the
additional flux at even larger radii than the larger scalable aperture
based on the encircled energy in the F210M PSF (typically a $\sim$0.15
mag correction).

\section{Source Selections}

\begin{table*}
 \setlength{\tabcolsep}{4pt}
\centering
\caption{Selection of $z\geq 8$ Galaxies by leveraging new {\it JWST}/NIRCAM Medium Band Data over the HUDF/XDF.}
\label{tab:cursample}
\begin{tabular}{c|c|c|c|c|c|c|c|c|c|c} \hline
     &    &     &            & & & & Lyman \\
     &    &     & \multicolumn{2}{c}{$z_{phot}$$^a$} & $M_{UV}$ & $m_{UV}$ & Break & $\Delta \chi^2$$^{a,d}$ &    &  \\
  ID & RA & DEC & EAzY & Prosp & [mag]$^b$ & [mag]$^b$ & [mag]$^c$ & & p($z$$>$5.5)$^a$ & Lit$^{1}$ \\\hline\hline
  \multicolumn{10}{c}{$z\sim8$-9 Selection}\\
XDFY-2376346017 & 03:32:37.64 & $-$27:46:01.7 & 8.3$_{-0.2}^{+0.2}$ & 7.9$_{-0.1}^{+0.1}$ & 
$-19.3$$\pm$0.1 & 28.0$\pm$0.1 & 1.4$\pm$0.2 & $-$45.0 & 1.000 & M13,B15\\
XDFY-2381345542 & 03:32:38.13 & $-$27:45:54.2 & 8.5$\pm$0.1 & 8.4$_{-0.0}^{+0.1}$ & $-19.6$$\pm$0.1 & 27.7$\pm$0.1 & 1.9$\pm$0.3 & $-$79.5 & 1.000 & B11,M13,O13,B15\\
XDFY-2394748078 & 03:32:39.47 & $-$27:48:07.9 & 8.5$\pm$0.1 & 8.4$_{-0.0}^{+0.1}$ & $-18.8$$\pm$0.1 & 28.5$\pm$0.1 & 1.5$\pm$0.8 & $-$35.3 & 1.000 & E13,M13,O13\\
XDFY-2392146324 & 03:32:39.21 & $-$27:46:32.5 & 8.6$_{-0.4}^{+0.3}$ & 8.2$_{-0.3}^{+1.1}$ & $-18.1$$\pm$0.2 & 29.2$\pm$0.2 & $>$1.9 & $-$15.9 & 1.000 & E13,M13,O13,B15\\
XDFY-2426447051 & 03:32:42.64 & $-$27:47:05.1 & 8.8$_{-0.3}^{+0.4}$ & 8.7$_{-0.3}^{+0.4}$ & $-18.4$$\pm$0.2 & 28.9$\pm$0.2 & $>$1.4 & $-$9.9 & 0.994 & E13,M13,O13\\
\\
\multicolumn{10}{c}{$z\sim10$-11 Selection}\\
XDFJ-2402448006 & 03:32:40.24 & $-$27:48:00.6 & 9.2$_{-0.6}^{+0.6}$ & 9.1$_{-0.4}^{+0.5}$ & $-17.9$$\pm$0.6 & 29.7$\pm$0.6 & $>$1.0 & $-$3.7 & 0.907 & O13,B15\\
XDFJ-2381146246$^{e}$ & 03:32:38.12 & $-$27:46:24.6 & 9.8$_{-0.6}^{+0.6}$ & 9.7$_{-0.4}^{+0.4}$ & $-18.1$$\pm$0.4 &  29.5$\pm$0.4 & $>$1.7 & $-$5.3 & 0.957 & B11,O13,B15\\
XDFJ-2404647339 & 03:32:40.47 & $-$27:47:33.9 & 11.4$_{-0.5}^{+0.4}$ & 10.8$_{-0.6}^{+0.6}$ & $-18.6$$\pm$0.2 & 28.9$\pm$0.2 & $>$1.5 & $-$10.0 & 0.993 & -- \\
\\
\multicolumn{10}{c}{$z\sim12$-13 Selection}\\
XDFH-2334046578 & 03:32:33.41 & $-$27:46:57.8 & 11.8$_{-0.5}^{+0.4}$ & 11.9$_{-0.6}^{+0.5}$ & $-18.6$$\pm$0.2 & 29.2$\pm$0.2 & 1.0$\pm$0.4 & $-$17.3 & 1.000 & -- \\
XDFH-2395446286$^f$ & 03:32:39.55 & $-$27:46:28.67 & 12.0$_{-0.2}^{+0.1}$ & 11.9$_{-0.1}^{+0.1}$ & $-20.0$$\pm$0.1 & 27.8$\pm$0.1 & 1.0$\pm$0.1 & $-$61.1 & 1.000 & B11,E13,M13,O13,\\
\multicolumn{10}{c}{} & B13\\
\hline\hline
\end{tabular}
\\\begin{flushleft}
$^1$ B11 = \citet{Bouwens2011_Nature}, E13 = \citet{Ellis2013}, M13 = \citet{McLure2013}, O13 = \citet{Oesch2013}, B15 = \citet{Bouwens2015_LF}\\
$^a$ Derived using EAzY  \citep{Brammer2008} and Prospector \citep{Johnson2021}\\
$^b$ Derived using the flux in the $H_{160}$, $HK_{182}$, and $K_{210}$ bands for sources in our $z\sim8$-9, $z\sim10$-11, and $z\sim12$-13 samples to probe the $UV$ luminosity at $\approx$1600\AA$\,$rest-frame. \\
$^c$ Amplitude of the nominal Lyman breaks for these $z\geq8$ galaxy candidates.\\
$^d$ $\chi^2_{\textrm{best},z>5.5} - \chi^2_{\textrm{best},z<5.5}$\\
\textbf{$^e$ $z_{\textrm{NIRSpec}}=10.38_{-0.06}^{+0.07}$ \citep{CurtisLake2022_JADESz13,Robertson2022_JADESz13} \\
$^f$ $z_{\textrm{NIRSpec}}=11.58\pm0.05$ \citep{CurtisLake2022_JADESz13,Robertson2022_JADESz13}}
\end{flushleft}
\end{table*}

\begin{figure*}
\centering
\includegraphics[width=2\columnwidth]{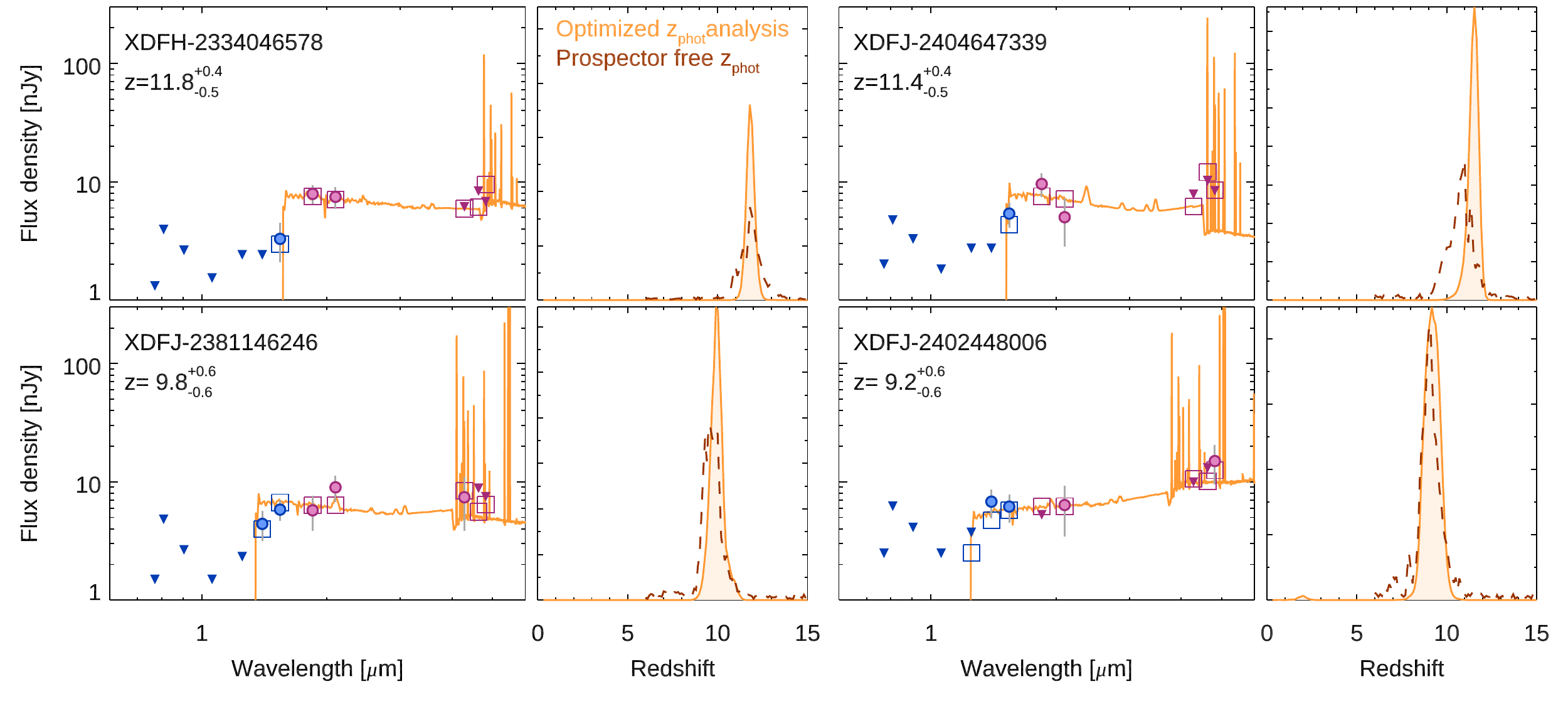}
\caption{Similar to Figure~\ref{fig:sedfit} but for the four high
  redshift candidates in our $z\sim10$-11 and $z\sim12$-13 samples.  In all
  four cases, the likelihood that the candidates are at very high
  redshift, i.e., $z>5.5$, is $>$90\%.  XDFJ-2381146246 has now been spectroscopically confirmed to have a redshift of 10.38$_{-0.06}^{+0.07}$ by the JADES team \citep{CurtisLake2022_JADESz13}.\label{fig:sedfit2}}
\end{figure*}

\subsection{Lyman Break Selections\label{sec:lbgselect}}

As in our previous searches for distant galaxies, we base our
selection on the use of two-color Lyman-break-like
criteria.  The first color in a Lyman-break selection ensures that candidate Lyman breaks in sources exceed some minimum amplitude (typically at least 1 mag) while the second color ensures the intrinsic colors of selected sources are blue.  Spectroscopic follow-up of sources identified with
two-color Lyman-break criteria have been shown to largely have the
redshifts that were targeted by the selection
\citep[e.g.][]{Steidel1999,Steidel2003,Stark2010,Ono2012,Finkelstein2013,Oesch2015,Zitrin2015,Oesch2016,Hashimoto2018_z9,Jiang2021}, and so it makes sense to continue using those criteria for even higher redshift selections with {\it JWST}.

In detail, we adopt the following two color criteria:
\begin{eqnarray*}
  ((Y_{105} - JH_{140} > 1.3) \vee \\
 ((Y_{105} - JH_{140} > 0.8) \wedge (J_{125}-H_{160} > 0.3)))\wedge \\
 (J_{125} - H_{160} < 0.9)
\end{eqnarray*}
for our $z\sim8$-9 selection,
\begin{eqnarray*}
  (J_{125}-H_{160}>0.9)\wedge (H_{160} - HK_{182} < 0.8) \wedge \\
  (H_{160}-K_{210}<0.8)
\end{eqnarray*}
for our $z\sim10$-11 selection, 
\begin{eqnarray*}
  (H_{160}-HK_{182}>0.7)\wedge (HK_{182} - K_{210} < 0.4)
\end{eqnarray*}
for our $z\sim12$-13 selection, and
\begin{eqnarray*}
  (HK_{182}-K_{210}>0.4)\wedge (K_{210} - F430M < 1.7)
\end{eqnarray*}
for our $z\sim14$-15 selection.  In applying the above color criteria, sources which are undetected in a band are set to their $1\sigma$ upper limits.  These color criteria are illustrated in Figure~\ref{fig:lbgcrit} relative to the expected colors of star-forming galaxies at $z>7$ and various low-redshift interlopers.

\begin{figure*}
\centering
\includegraphics[width=2\columnwidth]{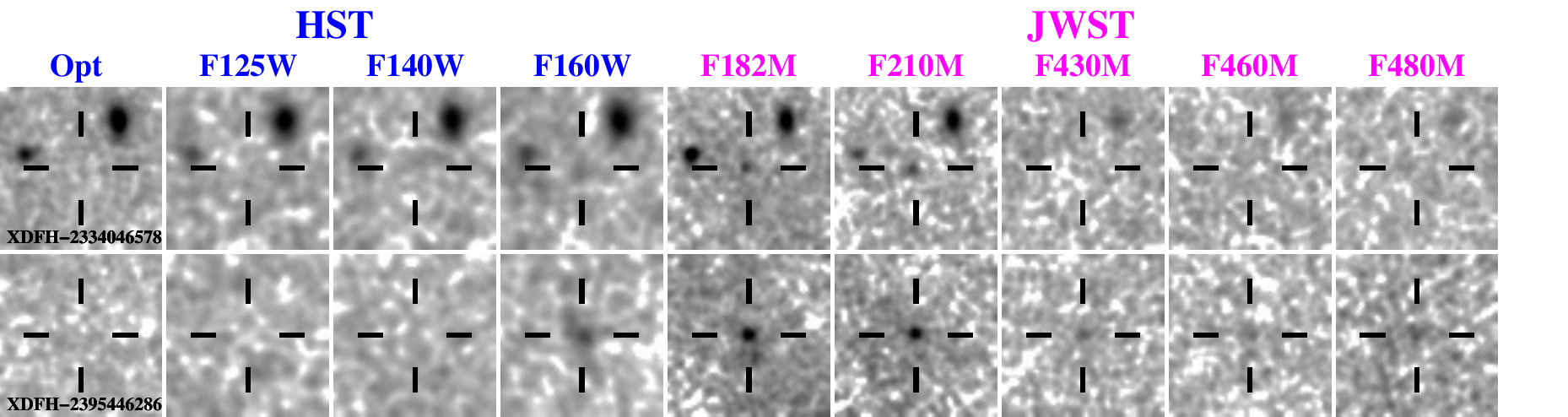}
\caption{Postage stamp images (2.4$''$$\times$2.4$''$) of the two
  $z\sim12$ galaxy candidates we identified over the HUDF using the
  new {\it JWST} medium band F182M, F210M, F430M, F460M, and F480M
  data and archival {\it HST} XDF observations in the F125W, F140W,
  and F160W bands.  The ``Opt'' column shows a stack of the
  observations in the {\it HST} F435W, F606W, F775W, F814W, F850LP,
  and F105W bands.  Both candidates show $\geq 4$$\sigma$ detections
  in both the F182M and F210M bands, but no detection in the F140W band or
  any bluer band.  XDFH-2395446286 is detected at even higher
  significance ($\geq$ 7.5$\sigma$) in the F182M and F210M bands,
  while also being well detected ($\gtrsim$4$\sigma$) in the redder
  F430M, F460M, F480M medium bands.  XDFH-2395446286 was first
  identified by \citet{Bouwens2011_Nature} as a probable $z$$\geq$ 10
  galaxy and then later argued to lie at $z\sim12$ because
  of the source's showing no detection in the F140W band observations
  \citep{Ellis2013,McLure2013,Oesch2013,Bouwens2013} obtained by the HUDF12
  program \citep{Ellis2013}.\label{fig:stamph}}
\end{figure*}

\begin{figure*}
\centering
\includegraphics[width=2\columnwidth]{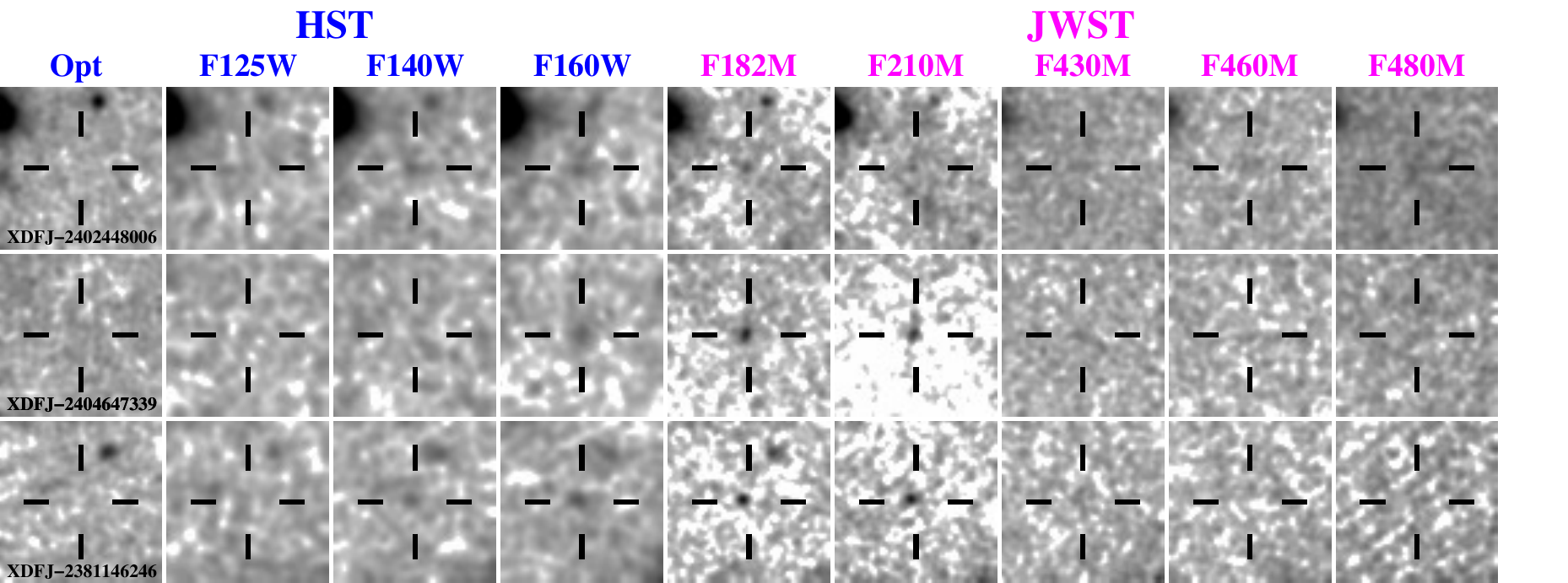}
\caption{Similar to Figure~\ref{fig:stamph} but for the three
  $z\sim10$-11 galaxies in our selection.  XDFJ-2381146246 was first
  identified as a candidate $z\sim10$ galaxy by
  \citet{Bouwens2011_Nature} and later confirmed as a strong candidate
  by \citet{Oesch2013} and \citet{Bouwens2015_LF}.  XDFJ-2402448006
  was first identified by \citet{Oesch2013} and again identified by
  \citet{Bouwens2015_LF}.  XDFJ-2404647339 is identified for the first
  time as a candidate $z\sim10$-11 galaxy here.}
    \label{fig:stampj}
\end{figure*}

To exclude lower-redshift interlopers, we coadd the flux from the
imaging observations blueward of the Lyman break using the $\chi^2$
statistic \citep{Bouwens2011_LF}, which we define as $\chi ^2 =
\Sigma_{i} \textrm{SGN}(f_{i}) (f_{i}/\sigma_{i})^2$ where $f_{i}$ is
the flux in band $i$ in a consistent aperture, $\sigma_i$ is the
uncertainty in this flux, and SGN($f_{i}$) is equal to 1 if $f_{i}>0$
and $-1$ if $f_{i}<0$.  Sources which show $\chi^2 > 9$ coadding the
flux in the $B_{435}V_{606}i_{775}I_{814}z_{850}$,
$B_{435}V_{606}i_{775}I_{814}z_{850}Y_{105}$, $B_{435}V_{606}i_{775}I_{814}z_{850}Y_{105}J_{125}$, and $B_{435}V_{606}i_{775}I_{814}z_{850}Y_{105}J_{125}JH_{140}H_{160}$ bands in our $z\sim8$-9, $z\sim10$-11, $z\sim12$-13, and $z\sim14$-15 selections, respectively, are excluded due to their possibly being
lower redshift interlopers.  We impose this criterion in three
apertures, i.e., a 0.2$''$-diameter aperture, a 0.35$''$-diameter
aperture, and the small scalable Kron apertures we use to make color
measurements.  Additionally, we exclude any $z\gtrsim10$ sources which
show a $3\sigma$ detection in the $Y_{105}$ band due to this band's
providing the most sensitive imaging at $>$1$\mu$m, which nevertheless
probes blueward of the Lyman break.

To ensure that the sources we detect are real, sources are required to
show a $5\sigma$ detection in the combined F182M+F210M image in a
0.35$''$-diameter aperture, to show a $3\sigma$ detection in the F182M
and F210M image individually in a 0.35$''$-diameter aperture, to show
a 4$\sigma$ detection in the combined F182M+F210M image in a
0.2$''$-diameter aperture.  Sources in our $z\sim10$-11 selection are
required to show a $3\sigma$ detection in either the F140W+F160W image
or the F160W image.

Following our selection of candidate $z\sim8$-13 sources based on the
described two color criteria, we computed redshift likelihood
functions $P(z)$ for each source using the EAZY photometric redshift
code \citep{Brammer2008}.  In fitting the photometry of individual
sources, we made use of spectral templates from the EAZY\_v1.0 set and
Galaxy Evolutionary Synthesis Models \citep[GALEV:][]{Kotulla2009},
which includes nebular continuum and emission lines according to the
prescription provided in \citet{Anders2003}, a $0.2 Z_{\odot}$
metallicity, and scaled to a rest-frame EW for H$\alpha$ of 1300\AA.
$>$80\% of the integrated redshift likelihood, i.e., $P(z>5.5)>0.8$,
is required to be at $z>5.5$ for sources to be retained in our $z\geq
8$ selections.

All candidate $z\geq8$ galaxies are then subjected to a visual
inspection and any sources that correspond to apparent diffraction
spikes or due to an artificially high background in a region of the image are excluded by hand.

Finally, to investigate whether our selection might be contaminated by lower-mass stars, we examined the SExtractor \textsc{Stellarity} parameter derived in the $HK_{182}$ and $K_{210}$ bands for all of the sources which satisfied our other selection criteria.  Given the narrow PSF in the $HK_{182}$ and $K_{210}$ bands, with an effective FWHM of 0.06-0.07$"$, these data provide us with the best constraints on whether sources are point-like.  Of the sources still remaining in our selection, XDFJ-2381146246 and XDFY-2394748078 are the most consistent with being point-like, with \textsc{Stellarity} parameters in excess of 0.8 in one or both the $HK_{182}$ and $K_{210}$ bands.  Comparing the $\chi^2$ of SED fits to these sources with SED fits to lower mass stars from the SpeX library \citep{Burgasser2014}\footnote{http://pono.ucsd.edu/$\sim$adam/browndwarfs/spexprism/index.html}, we find significantly better fits ($\Delta\chi^2 > 4$) to high-redshift star-forming templates from EAzY than to the lower-mass star templates.

The approximate redshift selection windows for our three $z\geq 8$
selections over the HUDF are shown in Figure~\ref{fig:zsel} and are
estimated based on the selection volume simulations described in
\S\ref{sec:uvlfs}.

\subsection{Selection Results}

In total, we identify 5 sources satisfying our $z\sim8$-9 selection
criteria, 3 sources satisfying our $z\sim10$-11 criteria, 2 sources
satisfying our $z\sim12$-13 criteria, and 0 sources satisfying our $z\sim14$-15 criteria.  The results of our selection are summarized in Table~\ref{tab:cursample}.

Remarkably, the brightest source is in our $z\sim12$-13 selection, and it was
previously identified in the available {\it HST} data over the HUDF as
UDFj-39546284 \citep{Bouwens2011_Nature}.  While the initial redshift
estimate was $z\sim10.3$ based on data from the HUDF09 program
\citep{Bouwens2011_LF}, the non-detection of the source in the
$JH_{140}$ band from the HUDF12 program \citep{Ellis2013} resulted in
the photometric redshift estimate increasing to $z=11.9\pm0.1$
\citep{Ellis2013,McLure2013,Bouwens2013,Oesch2013}.  Given the implied
luminosity of the source and very high redshift, an alternate
interpretation of the source was that it was an extreme emission line
candidate at $z\sim2$ \citep{Ellis2013,Bouwens2013,Brammer2013}.
\citet{Brammer2013} even presented evidence for a tentative
2.7$\sigma$ line detection at $\sim$1.6$\mu$m based on the available
WFC3/IR grism data over the HUDF.

Thanks to the availability of very deep NIRCam observations over the
HUDF, we can determine that the galaxy is actually at $z\sim12$ based
on the detection of a flat $UV$-continuum redward of the apparent
Lyman break and possible Balmer break at $\sim$4.8$\mu$m
(Figure~\ref{fig:sedfit}).  Our confidence in this conclusion is substantially bolstered by the exceptional sensitivity of the archival optical + near-IR observations in multiple passbands blueward of the break.  The source UDFj-39546284 appears to have
been the most distant galaxy detected with {\it HST} during its more than 30 years of operation. 

Parallel spectroscopy of XDFH-239446286 by the JADES GTO team with NIRSpec find strong evidence for a sharp spectral break in the source at $\sim$1.53$\mu$m, demonstrating that the source is at 11.58$\pm$0.05 \citep{CurtisLake2022_JADESz13,Robertson2022_JADESz13}, lower than our photometric redshift estimate of $z\sim 12$ but consistent within 2$\sigma$.

The best-fit SEDs for the three sources in our $z\sim10$-11 sample and
the second source in our $z\sim12$-13 sample are presented in
Figure~\ref{fig:sedfit2}.  Postage stamps of the sources in our
$z\sim12$-13 and $z\sim10$-11 samples are presented in
Figure~\ref{fig:stamph} and \ref{fig:stampj}.  Similar to XDFH-239446286, spectroscopy by the JADES GTO team confirm a second source from our selection XDFJ-2381146246 as lying at 10.38$_{-0.06}^{+0.07}$, within 1$\sigma$ of our photometric redshift estimate.

Postage stamps and SEDs for the five sources which make it into our
$z\sim8$-9 selection are presented in Figure~\ref{fig:stampy} and
\ref{fig:sedfit3} of Appendix A.  Interestingly enough, two of the
sources XDFY-2381345542 and XDFY-2394748078 show evidence for very
high EW [OIII]+H$\beta$ line emission based on the prominent
detections in both the F460M and F480M bands.  The contribution of the
emission lines to both bands indicate that the lines likely lie in the
overlap region between the two medium bands, allowing us to obtain very
precise constraints on their photometric redshifts from the
photometry to $z=8.5\pm0.1$.  The impact that strong [OIII]+H$\beta$ line emission can have in improving photometric redshift constraints recalls similar improvements made to $z\sim7$ sources leveraging {\it Spitzer}/IRAC photometry
\citep{Smit2014,Smit2015,Zitrin2015,RobertsBorsani2016,Bouwens2022_REBELS}.  While it may seem fortuitous to find two galaxies from our selection in such a small redshift interval, the presence of one galaxy in that interval makes this more likely.  We
will discuss this further in \S\ref{sec:stellar_pops}.

\begin{figure*}
\centering
\includegraphics[width=2\columnwidth]{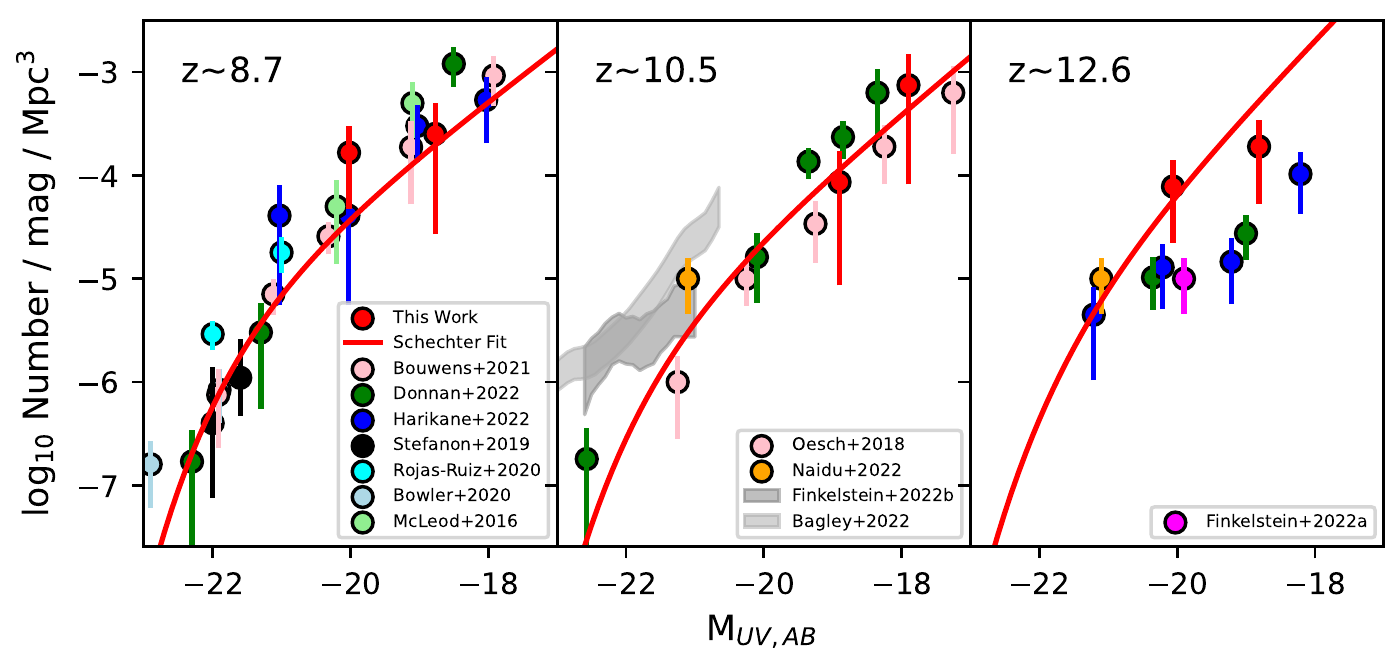}
\caption{Present determinations of the stepwise and Schechter $UV$ LF
  results at $z\sim8$-9, $z\sim10$-11, and $z\sim12$-13 (\textit{red circles}
  and \textit{red lines}, respectively) using the $z\geq 9$ candidate
  galaxies we have identified in the combined {\it HST}+{\it JWST}
  data set over HUDF.  Also shown are the LF results from \citet[][:
    pink circles]{Bouwens2021_LF}, \citet[][: green
    circles]{Donnan2022}, \citet[][: blue
    circles]{Harikane2022_z9to17}, \citet[][: black
    circles]{Stefanon2019}, \citet[][: cyan circles]{RojasRuiz2020},
  \citet[][: light blue circles]{Bowler2020_LF}, \citet[][: light green circles]{McLeod2016}, \citet[][: pink
    circles]{Oesch2018_z10LF}, \citet[][: dark grey-shaded region]{Finkelstein2022}, \citet[][: light grey-shaded region]{Bagley2022}, \citet[][: orange
    circles]{Naidu2022_z12}, and \citet[][: magenta
    circles]{Finkelstein2022_z12}.  Several other determinations of the $UV$ LF at $z\geq9$ \citep[][: which have been important historically]{McLure2013,Oesch2013,Oesch2014_z910LF,Bouwens2014_CLASH,Calvi2016,Bouwens2016_LF,Livermore2018,Morishita2018,Ishigaki2018,Bhatawdekar2019,Bouwens2019_LF,Morishita2021,Leethochawalit2022_HST} are not shown, but provide similar constraints to the others presented.  Our new LF results at $z\sim9$-10 are fairly similar to earlier work with {\it HST} and {\it JWST}.  While our $z\sim12$-13 LF estimates are higher than other recent determinations with {\it JWST}, we emphasize that the uncertainties in our results are large, being based on only two sources identified within a small volume.\label{fig:uvlfs}}
\end{figure*}

We also estimated photometric redshifts with \textsc{Prospector} \citep{Johnson2021}, adopting a flat redshift prior $0 < z_{phot} < 12$. \textsc{Prospector} runs on the \textsc{Flexible Stellar Population Synthesis (FSPS)} package (\citealt{Conroy2009, Conroy2010}) with the \textsc{Modules for Experiments in Stellar Astrophysics Isochrones and Stellar Tracks} (MIST; \citealt{Choi2016, Dotter2016}). For these measurements, we adopted a constant star-formation history (CSFH), a \citet{Chabrier2003} IMF defined between $0.1$ and $240 M_\odot$, a \citet{Calzetti2000} extinction curve, a  $Z_\mathrm{star}\equiv Z_\mathrm{gas}=0.2Z_\odot$ metallicity, and a ionization parameter  $\log U = -2.5$  (e.g., \citealt{Stark2017,deBarros2019}). We also adopted the nebular emission (both continuum and lines) estimates that natively come with the FSPS package (\citealt{Byler2017}), obtained by reprocessing the FSPS templates through \textsc{Cloudy} \citep{Ferland2013}. We refer the reader to \citet{Byler2017} for full details on the adopted procedure and for a detailed characterization of the results.

The agreement between the Prospector and EAzY results were excellent,
being within one sigma in all cases except one where the difference
was less than 2$\sigma$.  This agreement gives some confidence that
our photometric redshifts are fairly robust.\footnote{Strong emission
lines can significantly contribute to the flux density in those
photometric bands which intercept their emission. Because this
potentially introduces complementary constraints to the photometric
redshift measurements, we note that the templates corresponding to
ages of $3, 30$ and $300$ Myr have rest-frame equivalent widths EW$_0$
for [\ion{O}{2}]$_{\lambda3727}$ of
EW$_0$([\ion{O}{2}]$_{\lambda3727})\sim510, 170$, and $110$\AA\,
respectively, EW$_0(\mathrm{H}\beta)\sim590, 170, 80$\AA,
EW$_0$([\ion{O}{3}]$_{\lambda4959})\sim890, 120, 110$\AA, and
EW$_0$([\ion{O}{3}]$_{\lambda5007})\sim2700, 690, 330$\AA.}

All five of the sources in our $z\sim10$-11 and $z\sim12$-13 samples appear to be
very reliable, with $P(z>5.5)>0.9$ for all sources in the
selection and with 8 of the 10 showing $P(z>5.5)>0.99$.  This constitutes
a significantly higher fraction of reliable sources than we find \citep{Bouwens2022_EarlyNIRCam} / for the majority of the $z>8$
candidates reported over the SMACS0723 \citep{Pontop2022}, GLASS NIRCam parallel field \citep{Treu2022_GLASS},
and the first four pointings from CEERS.  The
higher reliability of the present selection of $z>8$ galaxies appears to be a consequence of the sensitivity of the archival observations
from {\it HST} over the field ($\geq$30.0 mag at $5\sigma$ from $\sim$0.4 to 0.8$\mu$m), being up to $\sim$2 mag deeper at optical wavelengths than observations over CEERS, SMACS0723, or the GLASS parallel.

We note the presence of a third possible $z\sim12$-13 candidate at
03:32:37.25, $-$27:46:01.9 that missed our selection due to the
detection of this source at $3\sigma$ in the $Y_{105}$ band in a
0.35$''$-diameter aperture.  While it is unclear how reliable that candidate
is based on archival imaging observations with {\it HST}, this should be easily
clarified using the very sensitive F115W observations the NIRCam GTO
team is obtaining as part of the JADES program \citep{JADES,Robertson2022}.

\subsection{Comparison with Earlier Selections with the {\it Hubble} Space Telescope and New Spectroscopic Sample from JADES}

Previously, \citet{Bouwens2011_LF}, \citet{Ellis2013},
\citet{McLure2013}, \citet{Oesch2013}, \citet{Bouwens2013}, and
\citet{Bouwens2015_LF} considered searches for $z\geq8.5$ galaxies
over the HUDF.  Of the sixteen unique $z\gtrsim 8.5$ candidate
galaxies reported as part of these analyses, our selection recovers
eight.

The remaining eight candidates do not satisfy our selection criteria
and there are a variety of reasons for this, which we detail in
Appendix B, but here we provide a brief summary.  For five of the
candidates, this is due to these candidates' failing to show 2$\sigma$
detections in the new medium-band F182M and F210M imaging observations with
NIRCam.  For another candidate (C5 from Appendix B), this is because the source is detected at
$>$5$\sigma$ significance in the $Y_{105}$ band.  For another candidate (C7 from Appendix B),
this is because the source was not identified in the SExtractor
catalog we constructed over the HUDF.  Finally, for the final remaining candidate (C6 from Appendix B), this
is because the integrated likelihood at $z>5.5$ does not exceed the 80\% threshold
required to be part of our $z\geq 8$ selection (\S\ref{sec:lbgselect}).

In parallel with the photometric identifcation of high-redshift
candidates conducted as part of our analysis, the JADES team also was
considering potential photometric candidate z$>$10 galaxies around the
HUDF.  NIRSpec follow-up observations conducted by the JADES succeeded
in confirming four of these candidates, finding redshifts of
10.38$_{-0.06}^{+0.07}$, 11.58$_{-0.05}^{+0.05}$,
12.63$_{-0.08}^{+0.24}$, and 13.20$_{-0.07}^{+0.04}$.  Of the four
sources, three lie inside the 4.6 arcmin$^2$ search field we consider
while one lies outside of the region.  Of the three sources in our
search field, two are found in our photometric sample.  The other does
not appear in our selection since it does not satisfy our
$H_{160}$-dropout selection criteria, due to its only showing a modest
spectral break ($\sim$0.4 mag) across the $H_{160}$ and $HK_{182}$
bands.

\section{Results}

\subsection{Luminosity Function Estimates\label{sec:uvlfs}}

The optical plus near-IR observations from HST over the HUDF and similarly deep medium-band NIRCam imaging data from JWST’s JEMS together constitute a very deep dataset that provides stronger constraints to the blue of the Lyman-alpha break than many of the datasets that have been available from the early JWST observations, allowing us to identify some of the highest-quality high-redshift candidates known to date.  This arguably puts us in position to obtain some of the most reliable $UV$ LF results to date in the early universe.

Given the small number of sources available over the HUDF, we derive
maximum likelihood results on the $UV$ LF at $z>8$ adopting Poissonian
statistics.  As in our own earlier analyses, we derive LF results by
maximizing the likelihood ${\cal L}$ of producing the observed
distribution of apparent magnitudes given some model LF:
\begin{equation}
{\cal L}=\Pi_i p(m_i)
\label{eq:ml}
\end{equation}
where the likelihood of a specific LF is computed over a set of
apparent magnitude intervals $m_i$.  Since Poissonian statistics are
assumed, the probability of finding $n_{\mathrm{observed},i}$
sources \begin{equation} p(m_i) = e^{-n_{\mathrm{expected},i}}
  \frac{(n_{\mathrm{expected},i})^{n_{\mathrm{observed},i}}}{(n_{\mathrm{observed},i})!}
\label{eq:mi}
\end{equation}
where $n_{\mathrm{observed},i}$ is the number of sources observed in a given
magnitude interval $i$ while $n_{\mathrm{expected},i}$ is the expected number
using some model LF.  The number of expected sources $n_{\mathrm{expected},i}$
is computed based on some model LF $\phi_j$ from the equation
\begin{equation}
n_{\mathrm{expected},i} = \Sigma _{j} \phi_j V_{i,j}
\label{eq:numcountg}
\end{equation}
where $V_{i,j}$ is the effective volume over which a source in the
magnitude interval $j$ can be both selected and have a measured
magnitude in the interval $i$.

\begin{table}
\centering
\caption{Binned LF Results for Galaxies at $z\geq8$}
\label{tab:binnedlfs}
\begin{tabular}{c|c}
\hline
$M_{UV}$ & $\phi^*$ [mag$^{-1}$ Mpc$^{-3}$]\\
\hline\hline
\multicolumn{2}{c}{$z\sim8$-9 galaxies}\\
$-$20.02 & 0.000166$\pm$0.000132\\
$-$18.77 & 0.000252$\pm$0.000250\\\\
\multicolumn{2}{c}{$z\sim10$-11 galaxies}\\
$-$18.90 & 0.000086$\pm$0.000120\\
$-$17.90 & 0.000746$\pm$0.000738\\\\
\multicolumn{2}{c}{$z\sim12$-13 galaxies}\\
$-$20.06 & 0.000078$\pm$0.000062\\
$-$18.81 & 0.000190$\pm$0.000152\\
\hline\hline
\end{tabular}
\end{table}

\begin{table}
\centering
\caption{Best-fit parameters derived for Schechter fits to the present $z\geq8$ $UV$ LF results}
\label{tab:lffits}
\begin{tabular}{c|c|c|c}
\hline
         & $\phi^*$ & $M^*$ & $\alpha$ \\
         & [$10^{-5}$ mag$^{-1}$ & & \\
Redshift &  Mpc$^{-3}$] & [mag] & \\
\hline
\multicolumn{4}{c}{Schechter}\\
8.7 & 1.5$_{-0.6}^{+1.0}$ & $-$21.15 (fixed) & $-$2.26 (fixed) \\
10.5 & 0.8$_{-0.4}^{+0.8}$ & $-$21.15 (fixed) & $-$2.38 (fixed) \\
12.6 & 1.6$_{-1.0}^{+2.0}$ & $-$21.15 (fixed) & $-$2.71 (fixed) \\
14.7 & $<$2.7$^a$ & $-$21.15 (fixed) & $-$2.93 (fixed)\\
\hline\hline
\end{tabular}
\\\begin{flushleft}
$^a$ Upper limit is $1\sigma$.
\end{flushleft}
\end{table}

\begin{table}
\centering
\caption{Inferred $UV$ Luminosity Densities$^1$ in the Early Universe Derived from the $z\geq8$ Search over the HUDF/XDF}
\label{tab:uvlumdens}
\begin{tabular}{c|c|c}
\hline
         & $\rho_{UV}$ & $\rho_{SFR}$ \\
Redshift & [ergs $s^{-1}$ Mpc$^{-3}$] & [$M_{\odot}$/yr/Mpc$^{3}$] \\
\hline\hline
8.7 & 25.15$\pm$0.24 & $-$3.00$\pm$0.24\\
10.5 & 24.33$\pm$0.30 & $-$3.82$\pm$0.30\\
12.6 & 24.91$_{-0.48}^{+0.37}$ & $-$3.24$_{-0.48}^{+0.37}$\\
14.7 & $<$25.56$^a$ & $<$$-2.59$$^a$\\
\hline\hline
\end{tabular}
\\\begin{flushleft}
$^1$ Luminosity densities integrated down to $-$18 mag.\\
$^a$ Upper limit is $1\sigma$.
\end{flushleft}
\end{table}

Selection volumes are computed for our four samples by inserting
artificial sources with various redshift and apparent magnitudes at
random positions within the NIRCam images and then attempting both to
detect the sources and to select them using our $z\sim8$-9, $z\sim10$-11,
$z\sim12$-13, and $z\sim14$-15 criteria.  We assume point-source sizes for
the galaxies we add to various images and assume the $UV$-continuum
slopes of sources to have a mean value of $-2.3$, with a $1\sigma$ scatter of 0.4.  The present size assumptions are not
especially different from that found for galaxies at $z\sim8$-17, both
using earlier HST observations \citep{Coe2013,Zitrin2014,Lam2019} and now using {\it JWST} 
observations \citep{Naidu2022_z12,Naidu2022_z17,Ono2022}.  These $UV$-continuum
slopes are in reasonable agreement with determinations available on
the basis of both {\it HST}+{\it Spitzer} data
\citep[e.g.,][]{Dunlop2013,Bouwens2014_betas,Wilkins2016,Stefanon2022_sSFR}
and now {\it JWST} data \citep{Topping2022_blueSlopes,Cullen2022}.

We computed selection volumes of $\sim$1750, $\sim$2500, $\sim$1500, and $\sim$1500 
cMpc$^3$ per arcmin$^2$ at brighter $UV$ luminosities for our $z\sim8$-9, 
$z\sim10$-11, $z\sim12$-13, and $z\sim14$-15 samples, respectively, corresponding 
to $\Delta z$ redshift slices of 1.1, 1.7, 1.0, and 1.0, respectively.  At fainter 
$UV$ luminosities near the limit of our selection, i.e., $-19$ to $-18$ mag, we 
estimated selection volumes of $\sim$1600, $\sim$2300, $\sim$870, and 
$\sim$30 cMpc$^3$ per arcmin$^2$.  While the correction factors can become quite large for faint sources in our highest-redshift selection, no sources are identified in those selections, so it only has a minimal impact on our LF results.

Recomputing selection volumes by applying the \citet{Ono2022}
$(1+z)^{-1.2}$ size scaling to the $z\sim 4$ galaxy population from
the HUDF \citep{Bouwens2015_LF}, we estimate selection volumes just
10\% lower than what we estimate adopting point-source sizes.  This
indicates that the uncertain sizes of higher redshift galaxies are
unlikely to have affected our results in any significant way.
Furthermore, it suggests it is improbable that there are many extended
sources at $z>8$ that have evaded our selections due to their large
sizes.

We derive determinations of the $UV$ LF results in 0.5-mag bins
and determinations of the best-fit Schechter function.  For the
Schechter function results, we fix the $M^*$ to $-21.15$ mag
consistent with the $z\geq 7$ $UV$ LF derived by
\citet{Bouwens2021_LF}, while we fix $\alpha$ to $-2.26$, $-2.38$, $-2.71$, and $-2.93$ at $z\sim8$-9, $z\sim10$-11, $z\sim12$-13, and $z\sim14$-15 consistent with an
extrapolation of the LF fit results of \citet{Bouwens2021_LF} to the
respective redshifts.  

We present our binned $UV$ LF results at $z\sim8$-9, $z\sim10$-11, and
$z\sim12$-13 LF results in both Table~\ref{tab:binnedlfs} and
Figure~\ref{fig:uvlfs}.  The parameterized fit results are presented
in Table~\ref{tab:lffits} and shown in Figure~\ref{fig:uvlfs} as the
red lines.  Encouragingly enough, our new LF results at $z\sim8$-9 are
consistent with what we obtained from our earlier analyses leveraging
a comprehensive set of HST fields \citep{Bouwens2021_LF}.  Agreement
with earlier results is not especially surprising given the
significant overlap between the present selection and earlier
selections.

At $z\sim10.5$, our results are consistent with what we previously
obtained in \citet{Bouwens2015_LF} and \citet{Oesch2018_z10LF}.  This is not especially surprising since both of these previous studies made use of the $z\sim10$ searches over the HUDF/XDF to constrain the faint end of the $UV$ LF and we use the same volume here, albeit with slightly better sampling of the volume just above $z\sim10$.

\begin{figure*}
\centering \includegraphics[width=2\columnwidth]{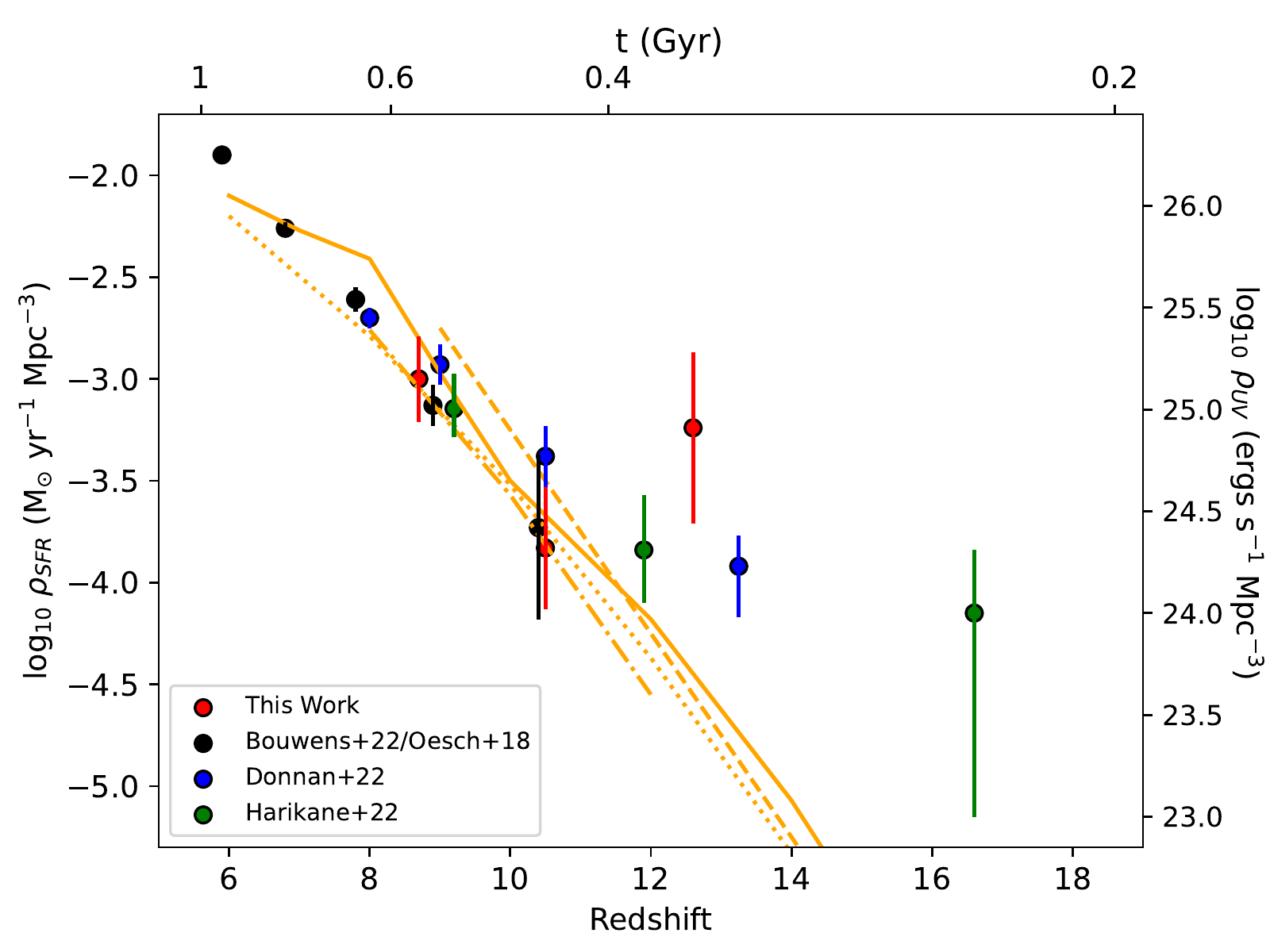}
\caption{$UV$ luminosity and star formation rate density integrated to
  $-18$ mag.  Shown are determinations from the present analysis
  (\textit{red circles}), \citet[][: \textit{blue
      circles}]{Donnan2022}, \citet[][: \textit{green
      circles}]{Harikane2022_LF}, and
  \citet{Bouwens2021_LF}/\citet[][: \textit{black
      circles}]{Oesch2018_z10LF}.  The orange lines indicate the
  expected evolution in the $UV$ luminosity density assuming no
  evolution in the star formation efficiency of galaxies across cosmic
  time using the models of \citet[][: solid]{Mason2015}, \citet[][:
    dot-dashed]{Tacchella2018}, \citet[][: dotted]{Bouwens2021_LF},
  and \citet[][: dashed]{Harikane2022_LF}.  The present determination
  of both the $UV$ luminosity and SFR density at $z\sim12$-13 is
  substantially higher than both other LF results in the literature,
  but is good agreement with earlier results at lower redshifts.  It
  is unclear why the $UV$ luminosity densities we derive at $z\sim12$
  are so much higher than other results, but we emphasize the present estimate is very uncertain, being based on just two sources identified within a small volume.  Whatever the case, the present LF results suggest that the $UV$ luminosity and SFR density may undergo a much milder evolution from early times than expected in many theoretical models.\label{fig:sfz}}
\end{figure*}

At $z\sim12.6$, our new LF results appear to be almost an order of
magnitude higher than found in the analyses of \citet{Donnan2022},
\citet{Harikane2022_z9to17}, and \citet{Finkelstein2022}, but
consistent with what \citet{Naidu2022_z12} find.  The substantially
higher normalization of our $z\sim12$-13 LF results follows from our
discovery of two $z\sim12$-13 candidates over the relatively
small-area 4.6 arcmin$^2$ HUDF.  The HUDF covers a $\sim$10$\times$
smaller area than used in other contemporary studies with {\it JWST},
e.g., \citet{Donnan2022}, \citet{Harikane2022_z9to17}, and
\citet{Finkelstein2022}.  As a result, one could imagine that our
results are subject to a larger large scale structure (LSS) uncertainty than
results obtained over wider areas, but the following estimate suggests that uncertainty is nevertheless modest.  Using the \citet{Trenti2008}
cosmic variance
calculator,\footnote{\url{https://www.ph.unimelb.edu.au/cgi-bin/mtrenti/prova.cgi}}
we estimate a $\sim$52\% RMS uncertainty in the normalization of our
$UV$ LF results at $z\sim12.6$ from cosmic variance which is smaller
than the uncertainties from small number statistics.

\begin{figure*}
\centering
\includegraphics[width=2\columnwidth]{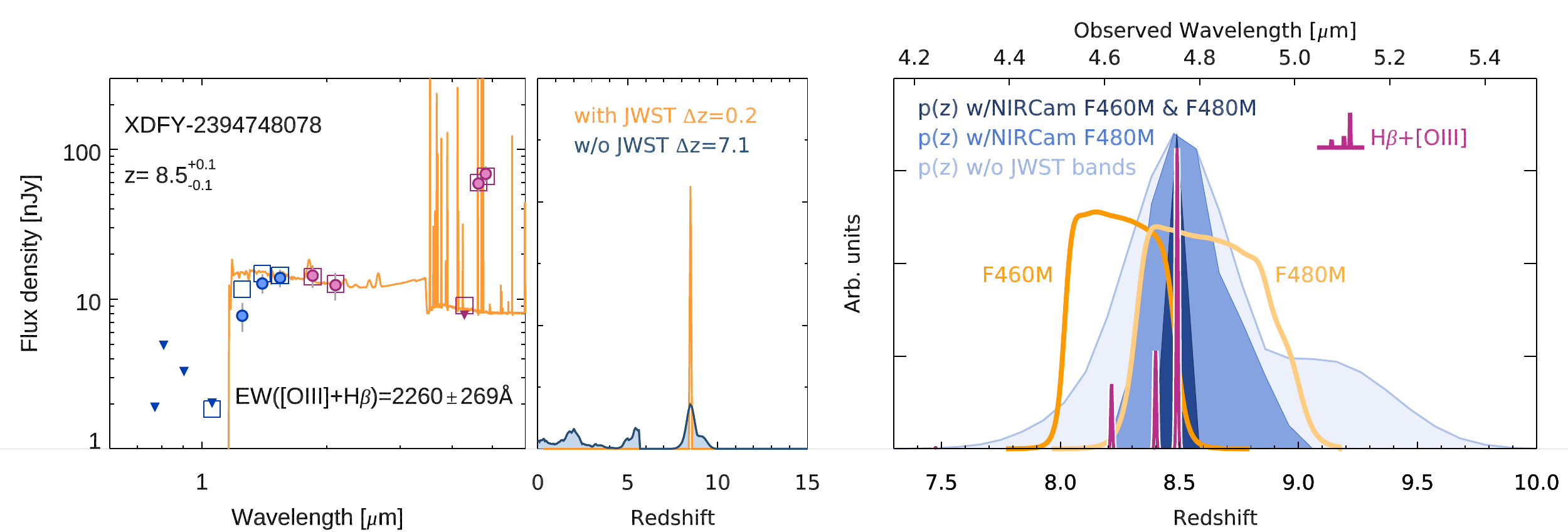}
\caption{(\textit{left}) Similar to Figure~\ref{fig:sedfit} but for one $z\sim8$-9 galaxy candidate appearing to contain prominent emission from the [OIII]+H$\beta$ lines in the F460M+F480M medium band filters demonstrating the dramatic impact the new JWST medium band observations have in improving the photometric redshift constraints on this candidate.  (\textit{right})  Illustration of the extremely precise constraints one can derive on the photometric redshift of sources when a prominent set of emission lines lie in the overlap region between two medium band filters F460M and F480M (whose sensitivity is shown with the orange lines).  Thanks to the prominent line emission in both bands, we can constrain the photometric redshifts of XDFY-2394748078 to
  $z=8.5\pm0.1$, which is much more precise than most of our other
  photometric redshift estimates and similar to the constraints we could obtain if these emission lines fell within one of NIRCam's narrow bands.  In addition, the medium band observations allow us to constrain the EWs
  [OIII]+H$\beta$ lines to 2260$\pm$269\AA.\label{fig:sedfit-pz}}
\end{figure*}

\subsection{Determinations of the Star Formation Rate Density}

With new determinations of the $UV$ LF leveraging the extremely deep
{\it JWST}+{\it HST} imaging observations over the HUDF, we are in
position to reexamine the evolution of the $UV$ luminosity density from early times.  When compared to other concurrent
analyses of JWST observations, results derived over the HUDF have the
advantage of being more robust against contamination from lower
redshift sources, but have the disadvantage of relying on a very small
volume, resulting in a larger uncertainty from cosmic variance.

We derive our constraints on the evolution of the $UV$ luminosity
density by directly integrating our stepwise $UV$ LF results to $-18$
mag -- given that this is the luminosity to which solid constraints exist.  We do not consider results fainter than this to avoid
extrapolating the results beyond where we can directly measure them from the observations.

We present our results in Table~\ref{tab:uvlumdens} and in
Figure~\ref{fig:sfz}.  We convert these results to a SFR density using
the conversion factor $\cal{K}$$_{FUV}$ is $0.7 \times 10^{-28}
M_{\odot}\,\textrm{year}^{-1} \,\textrm{erg}^{-1} \,\textrm{s}\,\textrm{Hz}$ from \citet{Madau2014}, which assumes a
\citet{Chabrier2003} IMF, a constant star formation rate, and
metallicity $Z=0.002$ $Z_{\odot}$.  To ensure that the results from
\citet{Oesch2018_z10LF}, \citet{Bouwens2021_LF}, \citet{Donnan2022}
and \citet{Harikane2022_z9to17} are presented to a similar limiting
luminosity, we have made a correction to several of new luminosity 
density results from {\it JWST} before
presenting them in Figure~\ref{fig:sfz}.  The correction was made by 
integrating the fiducial LFs from each of these studies down to both 
$-18$ mag (the limit we use) and $-17$ mag (the limit to which many 
results were quoted in the original studies) and then correcting results 
based on the difference.

To place the new results in context, we compare them to the results of
four different theoretical models which assume a constant star
formation efficiency.  These include \citet{Mason2015},
\citet{Tacchella2018}, \citet{Bouwens2021_LF}, and
\citet{Harikane2022_LF}.  Results from these models are shown on
Figure~\ref{fig:sfz} using the orange solid, orange dot-dashed, orange
dotted, and orange dashed lines, respectively.

Similar to other early studies of the evolution of the $UV$ LF from
$z\sim15$ \citep[e.g.][]{Donnan2022,Harikane2022_z9to17}, the present
results lie above the predictions of the constant star formation
efficiency models, suggesting that either star formation in the
earliest generation of galaxies is either substantially more efficient
than at later times, shows a significantly greater deviation relative
to the mean main-sequence evolution for galaxies \citep{Mason2022}, or
the mass-to-light ratio of stars is much lower, possibly indicative of
significantly modified IMF \citep{Steinhardt2022,
  Harikane2022_z9to17,Inayoshi2022}, or a number of other explanations \citep[e.g.][]{Ferrara2022_JWSTmodel,Mirocha2022_modelJWST,Harikane2022_z9to17,Kannan2022}.
  
While each of these possibilities would be exciting, it is clear that further study, including validation and investigation of these results with spectroscopy from {\it JWST}, and theoretical modeling \citep[e.g.][]{Mason2022,Wilkins2022,Kannan2022,BoylanKolchin2022,Lovell2022,Mirocha2022_modelJWST} will be required to ascertain what is driving the presence of so many luminous galaxies in the early universe.

\begin{table*}
\centering
\caption{Main stellar population parameters for our sample.}
\label{tab:stellar_pop}
\begin{tabular}{l|c|c|c|c|c|c|c|c|c|c} \hline
ID & $z_\mathrm{phot}^{a}$ & UV slope  & SFR$_\mathrm{UV}$$^c$ & SFR$_\mathrm{100}$$^d$ & $\log_{10}$ & sSFR$^{f}$ & $\log_{10}$ & $\log_{10}$ & $A_V$ & EW   \\
 &  & $\beta$$^b$ & & & $(M_\star/M_\odot)$$^e$ & & ($t_{50}$/yr)$^g$  &  ($t_{80}$/yr)$^g$ & & ([O\,III]+H$\beta$)$^h$ \\
 & & & [$M_\odot$/yr] & [$M_\odot$/yr] & & [Gyr$^{-1}$] & & & [mag] & [\AA]\\\hline\hline
\multicolumn{10}{c}{$z\sim12$-13 Selection} \\
\hline
XDFH-2395446286 & $12.0_{-0.2}^{+0.1} $ & $ -2.8_{-0.4}^{+1.1} $ & $  3.1_{- 0.5}^{+ 0.6} $ & $  3.3_{- 2.2}^{+ 3.0} $ & $  8.7_{- 0.5}^{+ 0.3} {**} $ & $   9.9_{-  7.0}^{+ 19.5} {**} $ & $  7.83_{-0.08}^{+0.07} {**} $ & $  7.99_{-0.05}^{+0.14} {**} $ & $ 0.08_{-0.05}^{+0.09} $ & $  \dots\dagger$ \\
XDFH-2334046578 & $11.8_{-0.5}^{+0.4} $ & $ -2.6_{-0.6}^{+1.8} $ & $  0.8_{- 0.2}^{+ 0.3} $ & $  1.0_{- 0.7}^{+ 1.3} $ & $  8.0_{- 0.6}^{+ 0.4} {**} $ & $  17.5_{- 14.8}^{+ 59.9} {**} $ & $  7.81_{-0.13}^{+0.11} {**} $ & $  8.01_{-0.07}^{+0.16} {**} $ & $ 0.10_{-0.07}^{+0.12} $ & $    \dots\dagger$ \\
\multicolumn{10}{c}{} \\
\multicolumn{10}{c}{$z\sim10$-11 Selection} \\
\hline
XDFJ-2404647339 & $11.4_{-0.5}^{+0.4} $ & $ -3.2_{-0.0}^{+0.8} $ & $  0.7_{- 0.2}^{+ 0.2} $ & $  0.9_{- 0.4}^{+ 1.5} $ & $  7.9_{- 0.6}^{+ 0.5} $ & $  24.1_{- 19.1}^{+ 86.4} $ & $  7.86_{-0.17}^{+0.19} $ & $  8.15_{-0.19}^{+0.10} $ & $ 0.18_{-0.12}^{+0.22} $ & $ \dots\dagger$ \\
XDFJ-2381146246 & $9.8_{-0.6}^{+0.6} $ & $ -1.6_{-0.9}^{+0.9} $ & $  0.5_{- 0.1}^{+ 0.1} $ & $  1.0_{- 0.6}^{+ 2.1} $ & $  7.4_{- 0.4}^{+ 0.5} $ & $  55.7_{- 41.4}^{+124.7} $ & $  7.49_{-0.48}^{+0.41} $ & $  8.16_{-0.28}^{+0.13} $ & $ 0.26_{-0.17}^{+0.27} $ & $  \dots\dagger$ \\
XDFJ-2402448006 & $ 9.2_{-0.6}^{+0.6} $ & $ -2.6_{-0.6}^{+1.3} $ & $  0.5_{- 0.1}^{+ 0.1} $ & $  1.6_{- 1.1}^{+ 2.5} $ & $  8.2_{- 0.6}^{+ 0.4} $ & $  19.0_{- 14.6}^{+ 54.5} $ & $  7.98_{-0.23}^{+0.25} $ & $  8.36_{-0.31}^{+0.08} $ & $ 0.62_{-0.36}^{+0.40} $ & $   \dots\dagger$ \\
\multicolumn{10}{c}{} \\
\multicolumn{10}{c}{$z\sim8$-9 Selection} \\
\hline
XDFY-2426447051 & $ 8.8_{-0.3}^{+0.4} $ & $ -2.9_{-0.2}^{+0.9} $ & $  0.6_{- 0.1}^{+ 0.1} $ & $  1.3_{- 0.9}^{+ 1.9} $ & $  7.9_{- 0.5}^{+ 0.4} $ & $  28.9_{- 22.8}^{+ 79.9} $ & $  7.74_{-0.27}^{+0.16} $ & $  8.07_{-0.16}^{+0.27} $ & $ 0.08_{-0.05}^{+0.09} $ & $  317\pm   139$ \\
XDFY-2392146324 & $ 8.6_{-0.4}^{+0.3} $ & $ -3.2_{-0.0}^{+1.0} $ & $  0.4_{- 0.1}^{+ 0.1} $ & $  0.7_{- 0.5}^{+ 1.0} $ & $  7.7_{- 0.6}^{+ 0.4} $ & $  25.0_{- 20.3}^{+ 80.6} $ & $  7.77_{-0.21}^{+0.14} $ & $  8.04_{-0.12}^{+0.31} $ & $ 0.07_{-0.05}^{+0.08} $ & $  <396$ \\
%XDFY-2404547501 & $ 8.6_{-7.9}^{+0.0} $ & $ -1.8_{-1.1}^{+1.1} $ & $  0.2_{- 0.0}^{+ 0.0} $ & $  0.1_{- 0.0}^{+ 0.1} $ & $  7.7_{- 0.4}^{+ 0.2} $ & $   2.4_{-  1.5}^{+  3.8} $ & $  7.92_{-0.05}^{+0.10} $ & $  8.23_{-0.25}^{+0.18} $ & $ 0.14_{-0.10}^{+0.15} $ & $  -15\pm   -15$ \\
XDFY-2381345542 & $ 8.5_{-0.1}^{+0.1} $ & $ -2.8_{-0.4}^{+0.4} $ & $  1.6_{- 0.2}^{+ 0.2} $ & $  1.7_{- 0.2}^{+ 0.4} $ & $  7.4_{- 0.1}^{+ 0.1} $ & $  75.8_{- 15.3}^{+ 18.3} $ & $  2.29_{-1.25}^{+0.96} $ & $  5.99_{-0.36}^{+0.27} $ & $ 0.01_{-0.01}^{+0.01} $ & $ 2313\pm   275$ \\
XDFY-2394748078 & $ 8.5_{-0.1}^{+0.1} $ & $ -1.8_{-0.4}^{+0.4} $ & $  0.8_{- 0.1}^{+ 0.1} $ & $  1.2_{- 0.6}^{+ 1.8} $ & $  7.3_{- 0.3}^{+ 0.4} $ & $  76.8_{- 49.9}^{+118.7} $ & $  6.64_{-1.42}^{+0.76} $ & $  8.05_{-0.54}^{+0.25} $ & $ 0.14_{-0.09}^{+0.14} $ & $ 2260\pm   269$ \\
%XDFY-2421347248 & $ 8.4_{-8.0}^{+***} $ & $ -1.9_{-0.9}^{+0.8} $ & $  0.5_{- 0.1}^{+ 0.1} $ & $  0.1_{- 0.1}^{+ 0.0} $ & $  7.9_{- 0.5}^{+ 0.2} $ & $   1.5_{-  1.1}^{+  2.7} $ & $  7.90_{-0.03}^{+0.04} $ & $  7.99_{-0.03}^{+0.28} $ & $ 0.06_{-0.04}^{+0.09} $ & $  185\pm   185$ \\
XDFY-2376346017 & $ 8.3_{-0.2}^{+0.2} $ & $ -2.7_{-0.4}^{+0.4} $ & $  1.4_{- 0.2}^{+ 0.2} $ & $  1.1_{- 0.7}^{+ 0.6} $ & $  8.3_{- 0.4}^{+ 0.2} $ & $   6.9_{-  4.3}^{+  9.3} $ & $  7.80_{-0.06}^{+0.04} $ & $  7.96_{-0.02}^{+0.02} $ & $ 0.03_{-0.02}^{+0.04} $ & $ \dots ^{*}$ \\
\hline
Median &  & $-2.7$ & $0.7$ & $1.1$ & $7.9$ & $24.5$ &  $7.80$ & $8.04$ & $0.09$ & $1288$\\
\end{tabular}
\\\begin{flushleft}
$^{a}$From EAzY \citep{Brammer2008} \\
$^b$UV slope, computed from the photometry in the bands probing rest-frame wavelengths between $1700$\AA\ and $2800$\AA.\\
$^c$Unobscured SFR measured from the rest-frame UV luminosity, following the relation of \citet{Madau2014} for a $Z=0.2Z_\odot$ metallicity.\\
$^d$Average SFR in over the last $100$\,Myr inferred from the total mass in stars formed over the last $100$\,Myr estimated by \textsc{Prospector} adopting the non-parametric SFH.\\
$^e$Mass in surviving stars. This was obtained by multiplying the total mass formed  since the onset of star formation by the fraction of surviving stars estimated by \textsc{Prospector} adopting the non-parametric SFH.\\
$^f$Specific star-formation rate, computed as the ratio between SFR$_{100}$ and $M_\star$.\\
$^g$ Lookback time, in logarithmic units, required to form the second half ($t_{50}$) and the last $80\%$ ($t_{80}$) of the stellar mass, inferred from the non-parametric SFH.\\
$^h$Cumulative equivalent width of H$\beta$ and [O\,III]$_{\lambda\lambda 4959,5007}$ emission lines, measured on the best-fit SED template. Values in parenthesis mark those measurements extracted for sources lacking photometric coverage in the bands where the H$\beta$ and [O\,III]$_{\lambda\lambda 4959,5007}$ line emissions is expected at the nominal redshift of the specific source.\\
$^{\dagger}$The available photometry does not provide coverage at wavelengths expected for H$\beta$ and [O\,III] line emission.\\
$^{\ddagger}$These measurements correspond to the photometric $2\sigma$ upper limits in the F480W band.\\
$^{*}$The continuum estimated by the SED fit is brighter than the $2\sigma$ upper limits in the bands where the H$\beta$ and [O\,III]$_{\lambda\lambda 4959,5007}$ line emissions is expected at the nominal redshift of the specific source.\\
$^{**}${These estimates exclusively rely on rest-UV light and are therefore potentially much more uncertain.}\\
\end{flushleft}
\end{table*}

\begin{figure*}
\centering
\includegraphics[width=2\columnwidth]{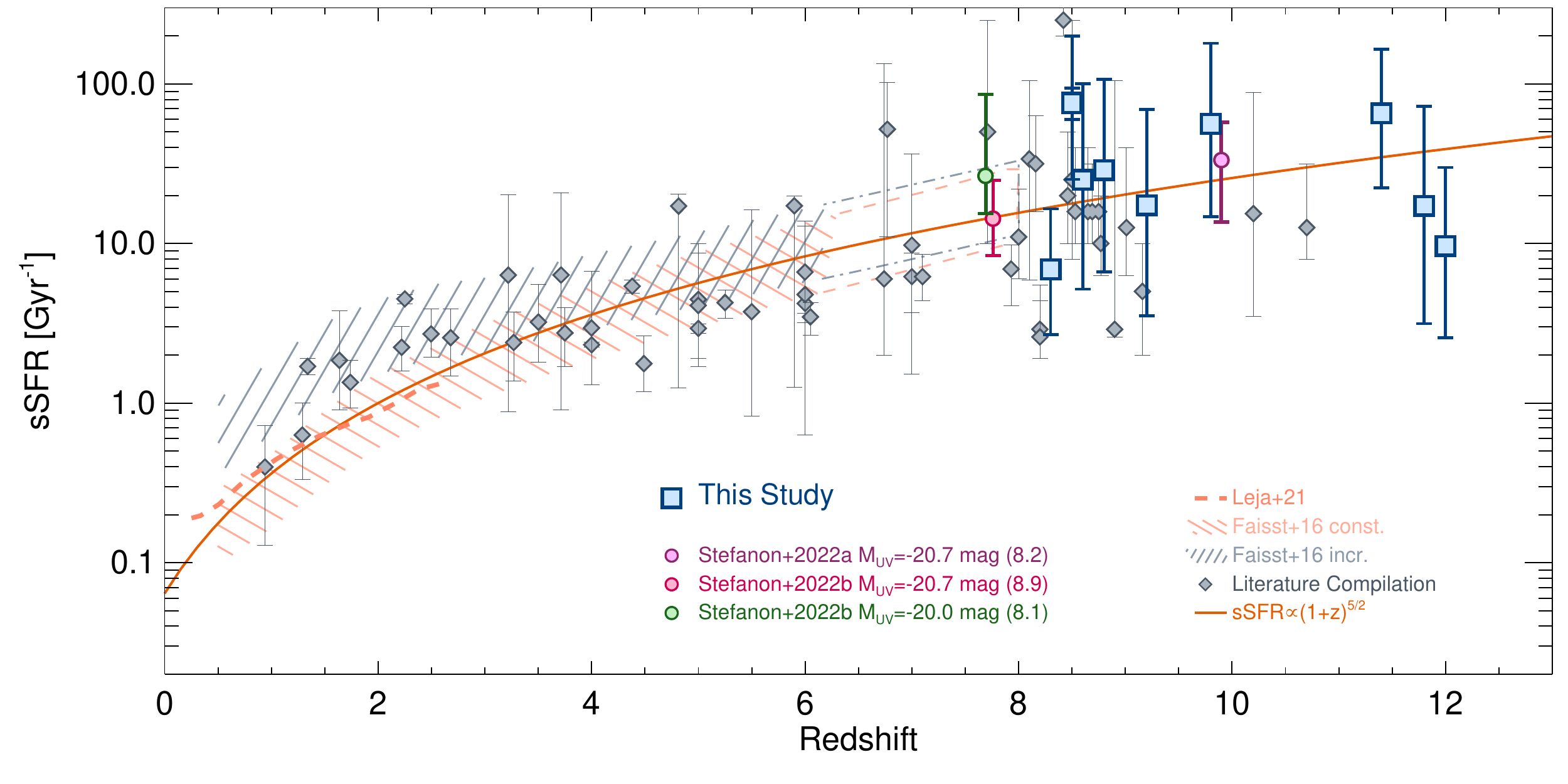}
\caption{Inferred sSFRs for faint $z\geq8$ galaxies identified over the HUDF/XDF (filled blue squares), and comparison to the evolution of the sSFR to lower redshift. The lower redshift estimates include the results of \citet{Stefanon2022_sSFR,Stefanon2022_IRACz10, Leja2022,Faisst2016}, and a compilation (solid grey diamonds) with the measurements from \citet{Labbe2013,Smit2014,Duncan2014,Salmon2015,Marmol2016,Santini2017,Davidzon2018,Khusanova2020,Strait2020,Endsley2021,Topping2022_REBELS,Tacchella2022,RobertsBorsani2022} and \citet{Bradley2022}. The orange solid curve corresponds to the model of \citet{Dekel2013}, which star formation is taken to be the direct result of the inflow of cold gas driven by the hierarchical assembly of the dark matter halos, and is modulated by a non-evolving efficiency of star formation. Our new measurements extend constraints on the evolution of the sSFR to $z\sim 12$.  Importantly, the results seem to remain  consistent with the toy model of \citet{Dekel2013}, suggesting that a marginally evolving star-formation efficiency is plausible starting as early as just $\sim370$\, Myr after the Big Bang.}
    \label{fig:ssfr}
\end{figure*}

\subsection{Stellar Population Analysis\label{sec:stellar_pops}}

Thanks to the depth of the {\it JWST} NIRCam observations at
$\geq2$$\mu$m and medium band coverage which allow for a probe of the
EW of emission lines like [OIII]+H$\beta$, we can place much tighter
constraints on the stellar population parameters of the faint $z\geq
8$ galaxies identified within the HUDF than was possible with {\it
  HST} (see also \citealt{Whitler2022_z10}, \citealt{Leethochawalit2022}, 
  \citealt{Furtak2022_jwst}, \citealt{Bradley2022} and 
  \citealt{Endsley2022_JWST}).

Table \ref{tab:stellar_pop} summarizes the main stellar population
parameters of our sample. These were estimated using the Bayesian tool
\textsc{Prospector} \citep{Johnson2021}, which runs on the
\textsc{Flexible Stellar Population Synthesis (FSPS)} package
(\citealt{Conroy2009, Conroy2010}) with the \textsc{Modules for
  Experiments in Stellar Astrophysics Isochrones and Stellar Tracks}
(MIST; \citealt{Choi2016, Dotter2016}). For our measurements, we
adopted a \citet{Chabrier2003} IMF defined between $0.1$ and $240
M_\odot$, a \citet{Calzetti2000} extinction curve, a
$Z_\mathrm{star}\equiv Z_\mathrm{gas}=0.2Z_\odot$ metallicity, a
ionization parameter $\log U = -2.5$ (e.g., \citealt{Stark2017,deBarros2019}), and a formation redshift of $z=20$
(\citealt{Mawatari2020,Hashimoto2018_z9, Tacchella2022,
  Harikane2022_LF}). We also adopted the nebular emission (both
continuum and lines) estimates that natively come with the FSPS
package (\citealt{Byler2017}), obtained by reprocessing the FSPS
templates through \textsc{Cloudy} \citep{Ferland2013}. We refer the
reader to \citet{Byler2017} for full details on the adopted procedure
and for a detailed characterization of the results.

Our estimates are based on a non-parametric SFH (e.g.,
\citealt{Leja2019b,Topping2022_REBELS,Whitler2022_Spitzer}), defined over
four bins in time measured backwards from the cosmic time
corresponding to the redshift of each source. Specifically, the four
bins were fixed at $0-3$Myr, $3-13$Myr, $13-100$ Myr, and $100-300$
Myr, respectively, with a Student's t-distribution continuity prior
modulating the ratio of the SFR in contiguous bins, with $\nu=2$ and
$\sigma=2$ (see \citealt{Stefanon2022_Halpha}, and discussions in
\citealt{Leja2019a, Tacchella2022}). This configuration can
concurrently accommodate a recent burst of star formation and
significant star formation during the initial assembly of the galaxy
(\citealt{Hashimoto2018_z9,RobertsBorsani2020,Tacchella2022,Whitler2022_Spitzer}). We note here that previous tests with up to 8 bins in time (0-3Myr,
3-13Myr, 13-100 Myr, and log-spaced afterwards) found marginal
differences in the final parameter estimates (e.g.,
\citealt{Stefanon2022_Halpha}). This is indeed reassuring, given the
coarse spectral resolution provided by the currently available
photometry can pose severe challenges to any attempts of constraining
the SFH of individual sources.

We present two distinct SFR measurements. One (SFR$_{100}$) is
obtained by averaging the total mass formed in the last $100$\,Myr as
estimated through the template analysis with \textsc{Prospector},
while the other (SFR$_\mathrm{UV}$) is computed by converting the UV
luminosity using the factors listed by \citet{Madau2014}, interpolated
for a $Z=0.2Z_\odot$ metallicity, a \citet{Chabrier2003} IMF, and adopting a negligible dust
emission, as recent results at high redshifts suggest (e.g.,
\citealt{Bouwens2016_ALMA, Dunlop2017, McLure2018, Bouwens2020}), and
consistent with the small $A_V$ values obtained with the
non-parametric analysis. We finally computed sSFR
($\equiv$SFR/$M_\star$) combining the SFR$_{100}$ and $M_\star$ values
previously estimated.

We computed UV slopes $\beta$ following the procedure of \citet{Stefanon2019}. Briefly, we constructed a grid of UV slope values based on stellar population models from the \citet{Bruzual2003} template set and attenuation $A_V=0$-$3$\,mag. We then fit the corresponding synthetic flux densities in the bands blue-ward of the rest-frame $2800$\,\AA\ to the observations. We computed uncertainties by randomly perturbing the observed flux densities and refitting. This procedure allows us to make full use of the near-IR data and to naturally take into account redshift uncertainties.

Remarkably, the observations in the medium bands provided by JEMS program \citep{Williams2023_JEMS} 
not only enable pinpointing the
photometric redshifts for three sources in our
sample. The $\sim3-10\sigma$ detections in the NIRCam/F460M and F480M
also allow for accurate estimates of the intensity of the H$\beta$ and
[O\,III]$_{\lambda\lambda4959,5007}$ lines for redshifts
$z\sim8.5$. We measured the equivalent widths (EW) of these lines from
the ratio between the flux density in the F480M band and the continuum
inferred from the best-fitting SED template, given the current lack of
$>2\sigma$ detection in bands probing the continuum free from emission
lines in our sample. Uncertainties in the EW were computed by randomly
perturbing the flux densities according to the corresponding $1\sigma$
uncertainties 1000 times, and repeating the measurement following our
main procedure. We finally verified that our EW measurements were
consistent at $\le 1\sigma$ with the cumulative
EW(H$\beta$+[O\,III]) measured on the best-fitting SED templates.\\

Overall, we find markedly blue UV slopes (median $\beta_\mathrm{med}\sim -2.7$), elevated sSFR (sSFR$_\mathrm{med}\sim 24.5$\,Gyr$^{-1}$) and extreme nebular line emission (EW$_\mathrm{med}$([O\,III]+H$\beta)\sim1300$\,\AA), suggesting very young stellar populations. Specifically, our UV slopes measurements are consistent with recent results at similar redshifts based on JWST data (e.g., \citealt{Whitler2022_z10, Bradley2022}), and consistent although marginally bluer than the measurements of \citet{Cullen2022}.

Perhaps unsurprisingly, the typical stellar masses characterizing our sample lie in range $\log{M_\star/M_\odot}\sim7.3$-8.7, populating the low-mass end of the stellar mass function at these redshifts (e.g., \citealt{Stefanon2022_sSFR}). The associated sSFR  range between sSFR$\sim7-80$\,Gyr$^{-1}$, consistent with recent measurements (e.g., \citealt{Bradley2022}), and implying mass assembly timescales of $\lesssim 10$-100$\,$Myr (for a constant SFH). Such rapid assemblies are also qualitatively supported by the typical formation times as estimated through the non-parametric SFH ($t_{50}$ and $t_{80}$ in Table \ref{tab:stellar_pop}). 

Our sSFR measurements are presented in Figure \ref{fig:ssfr}, together with previous estimates at $z\sim8$ by \citet{Stefanon2022_sSFR}, at $z\sim10$  by \citet{Stefanon2022_IRACz10}, and a compilation of measurements including \citet{Labbe2013,Smit2014, Duncan2014, Salmon2015,Faisst2016,Marmol2016, Santini2017, Davidzon2018, Khusanova2020,Strait2020,Endsley2021, Topping2022_REBELS,Tacchella2022,RobertsBorsani2022} and \citet{Bradley2022}. Our new estimates push current constraints on the sSFR to $z\sim12$ (although the stellar masses $M_\star$ of $z>10$ galaxies in our selection only have weak constraints on their rest-frame optical light from the F480M flux measurements); moreover, comparison with previous estimates further supports a scenario where the sSFR monotonically increases with increasing redshift (see e.g., \citealt{Stefanon2022_sSFR, Stefanon2022_IRACz10}) up to $z\sim12$. 

To put these new results in context, in Figure \ref{fig:ssfr} we compare the compilation of sSFR estimates with the predictions of the model by \citet{Dekel2013}. These predictions were obtained from analytical considerations on the Extended Press-Schechter formalism (see also \citealt{Neistein2008, Weinmann2011, Genel2014}), and assuming a non-evolving conversion factor between the specific accretion rate of the dark matter halos and the sSFR. The remarkable agreement between measurements out to $z\sim12$ and the model suggests a marginally evolving star formation efficiency out to just $\sim370$\,Myr into cosmic time.

Finally, the accurate EW measurements enabled by the NIRCam medium-band filters reveal extreme [O\,III]+H$\beta$ line emission, $\sim1.5\times$ more intense than previous estimates based on \textit{Spitzer}/IRAC colors (e.g., \citealt{Stefanon2022_sSFR}), and further support the scenario that galaxies with very intense radiation fields are ubiquitous in the early Universe (e.g., \citealt{Schaerer2022}).

\section{Summary}

Here we have taken advantage of sensitive medium-band observations
acquired over the {\it Hubble} Ultra Deep Field from the JEMS program \citep{Williams2023_JEMS} to extend searches for
star-forming galaxies there to even higher redshift, i.e., $z>10$, than was possible
with the {\it Hubble} Space Telescope.  The medium band observations are a
good match to the depth of the {\it HST} observations but probe redward to
2.1$\mu$m in the short wavelength channel, allowing for sensitive
searches for galaxies to $z\sim15$.

Searches for $z\geq 8$ star-forming galaxies over the {\it Hubble} Ultra
Deep Field have the important advantage of having the most sensitive
available supporting observations with {\it HST}, resulting in the
most reliable selections of $z\geq 8$ galaxies for which spectroscopy is not yet available.

Using the combined {\it HST} + {\it JWST} data set, we construct
sensitive photometric catalogs of all the sources over the {\it
  Hubble} Ultra Deep Field and employ a two-color Lyman-break
selection to search for star-forming galaxies at $z\sim8$-9,
$z\sim10$-11, $z\sim12$-13, and $z\sim14$-15.  The mean redshift estimated for these
selections using selection volume simulations is 8.7, 10.5, 12.6, and 14.7,
respectively.

Interestingly enough, based on these selections, we find that the
highest redshift candidate source identified in {\it HST} data, i.e.,
UDFj-39546284 \citep{Bouwens2011_Nature}, does in fact have a redshift of $z=12.0_{-0.2}^{+0.1}$, as suggested by analyses by \citet{Ellis2013},
\citet{McLure2013}, \citet{Oesch2013}, and \citet{Bouwens2013}.  Parallel spectroscopy from the JADES team \citep{CurtisLake2022_JADESz13,Robertson2022_JADESz13} find that the source has a spectroscopic redshift of 11.58$\pm$0.05.  The source is thus the most distant galaxy identified by {\it HST} in its more than 30 years of operation.  It is not an extreme emission line source
as had been alternatively suggested earlier by \citet{Ellis2013}, \citet{Bouwens2013}, and
\citet{Brammer2013}.

Besides this $z\sim12$ source, we find 5, 3, 1, and 0 additional galaxies
in our $z\sim8$-9, $z\sim10$-11, $z\sim12$-13, and $z\sim14$-15 selections, two of
which are entirely new discoveries from the {\it JWST} data.  The
photometric redshifts of the new sources are $z\sim11$ and $z\sim12$,
which is in excess of what is easily identified in the available {\it HST} data over the HUDF/XDF.

Leveraging these new selections, we run selection volume simulations
and compute $UV$ LF results at $z\sim8.7$, $z\sim10.5$, 
$z\sim12.6$, and $z\sim14.7$ (\S\ref{sec:uvlfs}).  At $z\sim8.7$ and $z\sim10.5$, our results are
consistent with earlier HUDF results at these redshifts.  However, at
$z\sim12.6$, we infer a 5-10$\times$ higher normalization for the $UV$
LF than for contemporary analyses in the literature and more
consistent with results at $z\sim8$-9.

Given the limited volume and the presence of only two sources in our
$z\sim12.6$ selection, our $UV$ LF results at $z\sim12.6$ are
accordingly very uncertain.  Nevertheless, like other early results on
the $UV$ LF evolution with {\it JWST} from $z\sim15$, our results lie
substantially in excess of the predictions of constant star formation
efficiency models that are effective in modeling the evolution of $UV$
LF and galaxy properties for much of cosmic time
\citep{Mason2015,Tacchella2018,Bouwens2021_LF,Harikane2022_LF}.  It is unclear
what this means, but could mean that star formation is substantially
more efficient than at later times, the evolution of the main sequence
of star formation in galaxies shows a much greater variation relative
to the mean \citep{Mason2022}, or the mass-to-light ratio of star
formation (and hence the IMF) is substantially shifted relative to
later times \citep{Steinhardt2022,Harikane2022_z9to17}.

We also made use of the NIRCam imaging redward of 2$\mu$m to characterize the stellar populations of the sources and rest-frame EWs of [O\,III]+H$\beta$ (\S\ref{sec:stellar_pops}).  Based on our analysis, we find that the typical star-forming galaxy in our sample had a very blue $UV$-continuum slope $\beta\sim-2.7$, an elevated sSFR $\sim$ 24.5 Gyr$^{-1}$, while showing extremely high-EW line emission in [O\,III]+H$\beta$ with a median EW of $\sim$1300\AA.

We note that while our $z\sim12.6$ $UV$ LF results are in excess of what one might predict for constant star formation efficiency models, the evolution we find for the sSFRs of galaxies from $z\sim12$ is nonetheless in better agreement (Figure~\ref{fig:ssfr}).  While it is unclear what would resolve these tensions, perhaps the small HUDF volume shows a particular overdensity of $z\sim12$ sources or there is an evolution in the IMF from high-redshift as speculated in multiple studies \citep[e.g.,][]{Steinhardt2022,
  Harikane2022_z9to17,Inayoshi2022}.  Clearly, this is an issue that will need to be investigated further in future data sets.

The presence of extreme line emission from [OIII]+H$\beta$ (EWs of $\sim$2300\AA) is particularly convincing in the cases of XDFY-2381345542 and
XDFY-2394748078 due to the substantial flux excesses seen in the F460M and F480M medium bands (Figure~\ref{fig:stampy}) over that seen in the F430M band.  The present inferences regarding the stellar populations of $z\sim8$-13 galaxies extend the trend in galaxy properties seen from lower redshifts to $z\geq8$ with {\it HST}+{\it Spitzer} \citep{Stefanon2022_sSFR} and also with {\it JWST} \citep{Endsley2022_JWST,Cullen2022,Bradley2022,Topping2022_blueSlopes,Furtak2022_jwst,Leethochawalit2022}.

In the future, deeper imaging with NIRCam and NIRSpec spectroscopy
should become available for many of the sources in our $z\geq 8$
selections from the JADES program, e.g., as in \citet{CurtisLake2022_JADESz13} and \citet{Robertson2022_JADESz13}.  This will allow for a more sensitive probe down the
$UV$ LFs at $z\geq 8$ and also a search for fainter sources to even
higher redshift.  Additionally, it should facilitate spectroscopic
confirmation and characterization of the faint $z\geq 8$ galaxies
located within the HUDF, providing essential insight into the sources
that likely reionized the universe
\citep[e.g.][]{Bouwens2015_Reion,Robertson2022}.

\section*{Acknowledgements}

We are appreciative to the co-PIs Christina Williams, Michael Maseda, and Sandro Tacchella of the JEMS program we utilize for their scientific vision and initiative in making possible sensitive medium band data over the Hubble Ultra Deep Field, which enabled the science described in this manuscript.  
The authors would like to thank Ivo Labb{\'e} for helpful feedback on the science described here.  RJB acknowledges support from NWO grants 600.065.140.11N211 (vrij
competitie) and TOP grant TOP1.16.057. MS acknowledges support from the CIDEGENT/2021/059 grant, from project PID2019-109592GB-I00/AEI/10.13039/501100011033 from the Spanish Ministerio de Ciencia e Innovaci\'on - Agencia Estatal de Investigaci\'on, and from Proyecto ASFAE/2022/025 del Ministerio de Ciencia y Innovación en el marco del Plan de Recuperación, Transformación y Resiliencia del Gobierno de España  RPN acknowledges funding from JWST programs GO-1933 and GO-2279. Support for this work was provided by NASA through the NASA Hubble Fellowship grant HST-HF2-51515.001-A awarded by the Space Telescope Science Institute, which is operated by the Association of Universities for Research in Astronomy, Incorporated, under NASA contract NAS5-26555. PAO acknowledges support from: the Swiss National Science Foundation through project grant 200020 207349.  The Cosmic Dawn Center (DAWN) is funded by the Danish National Research Foundation under grant No. 140.  This work is based on observations made with the NASA/ESA/CSA James Webb Space Telescope. The
data were obtained from the Mikulski Archive for Space Telescopes at the Space Telescope Science Institute, which is operated by the Association of Universities for Research in Astronomy, Inc., under NASA contract NAS
5-03127 for JWST. These observations are associated with {\it JWST} program \# 1963.

\section*{Data Availability}

All data used here are available from the Barbara A. Mikulski Archive for Space Telescopes (MAST: \url{https://mast.stsci.edu}), both in the form of raw and high level science products.

\bibliographystyle{mnras}
\bibliography{main} % if your bibtex file is called example.bib                                             
\appendix

\section{Postage stamps and SEDs for Current Sample of $z\sim8$-9 Galaxies}

\begin{figure*}
\centering
\includegraphics[width=2\columnwidth]{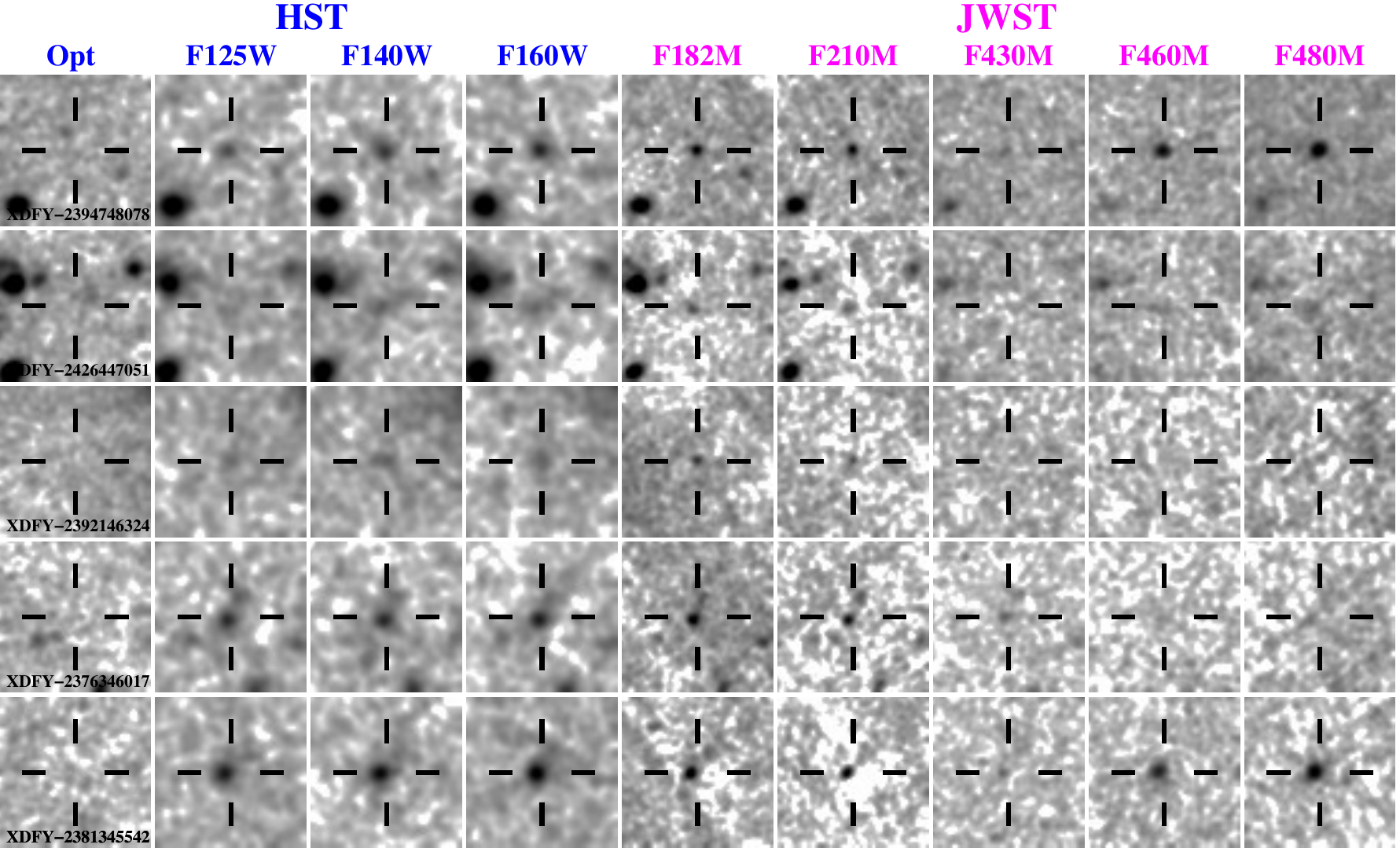}
\caption{Similar to Figure~\ref{fig:stamph} but for the five
  $z\sim8$-9 galaxy candidates in our selection.  All five candidate
  had been identified as part of earlier $z\sim8$-9 selections over
  the HUDF.  The particularly bright detections in the F460M and F480M
  bands for two sources from this selection, XDFY-2394648078 and
  XDFY2381345542, almost certainly arise due to line emission from
  [OIII]+H$\beta$, allowing us to place very tight constraints on
  their photometric redshifts as well as the EW of the [OIII]+H$\beta$
  lines (see Figure~\ref{fig:sedfit-pz}).\label{fig:stampy}}
\end{figure*}

\begin{figure*}
\centering
\includegraphics[width=2\columnwidth]{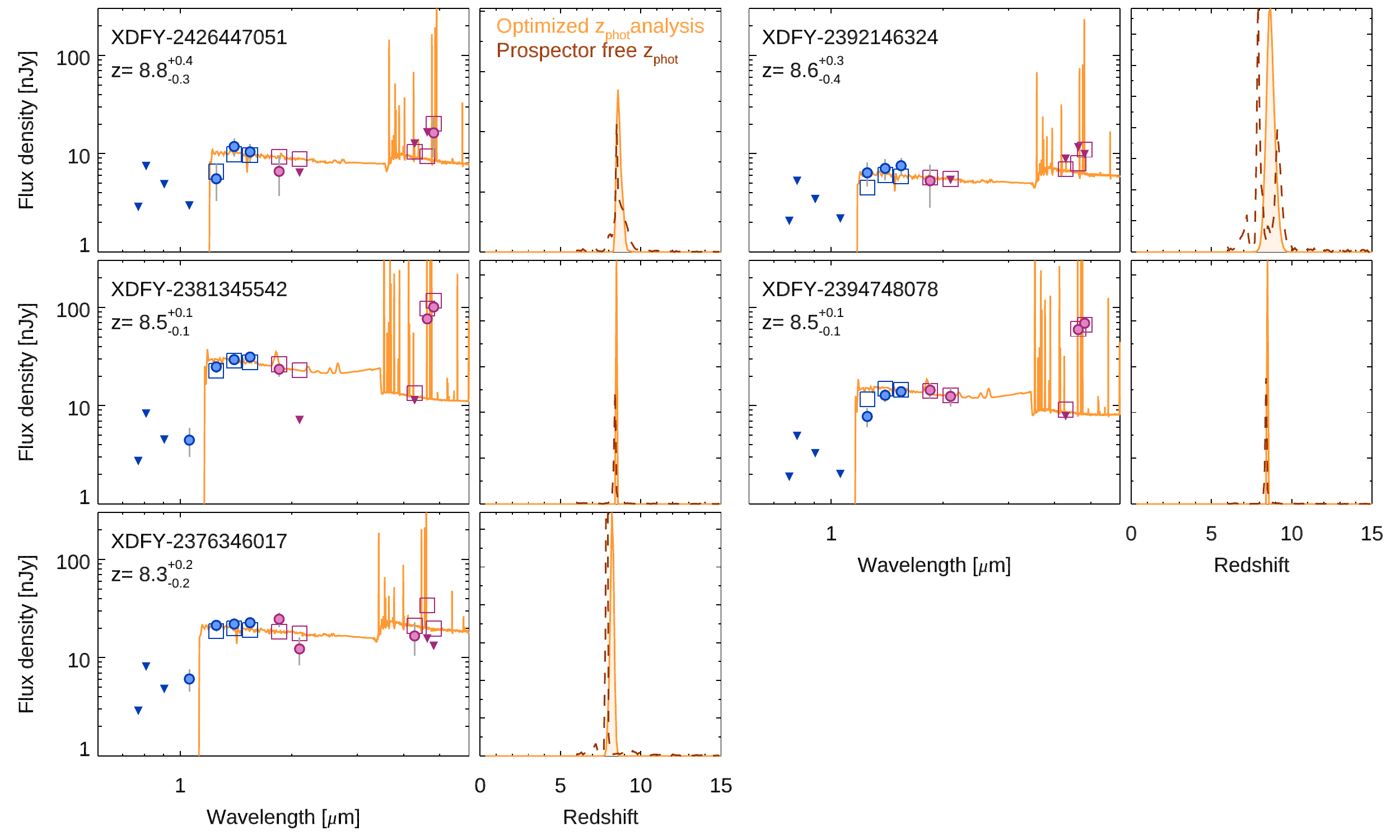}
\caption{Similar to Figure~\ref{fig:sedfit} but for galaxies in our
  $z\sim8$-9 sample.  Note the dramatic differences in the tightness of the redshift constraints on $z\sim8$-9 sources with prominent line emission contributing to the medium bands (e.g., XDFY-2381345542 and XDFY-2394748078) and those sources without such emission (e.g., XDFY-2392146324).\label{fig:sedfit3}}
\end{figure*}

Here we present the postage stamp images and model SEDs we derive for
the present selection of $z\sim8$-9 galaxies identified over the HUDF.
The postage stamp images are presented in Figure~\ref{fig:stampy}
while the best-fit model SEDs for these sources are presented in
Figure~\ref{fig:sedfit3}.

\section{Analysis of Various $z=8$-10 Candidate Galaxies Previously Reported in the Literature}

Here we consider various $z=8$-10 candidate galaxies which had
previously been reported in various analyses but did not meet the
current selection criteria.  These sources were derived from a number
of previous analyses of the HUDF including \citet{Bouwens2011_Nature},
\citet{Ellis2013}, \citet{McLure2013}, \citet{Oesch2013}, \citet{Bouwens2013}, and
\citet{Bouwens2015_LF}.  There are eight such sources, and they are
listed in Table~\ref{tab:litsample} with source IDs running from C1 to
C8.

For our assessment, we place 0.35$''$-diameter apertures at the
reported coordinate of the sources (after accounting for the 0.3$''$
shift in the coordinates of the sources to account for the improved
astrometry of the HUDF data) but otherwise compute total magnitudes
using scalable Kron apertures corrected to total using the estimated
encircled energy on the wings of the {\it JWST} F182M-band PSF.

\begin{figure*}
\centering
\includegraphics[width=2\columnwidth]{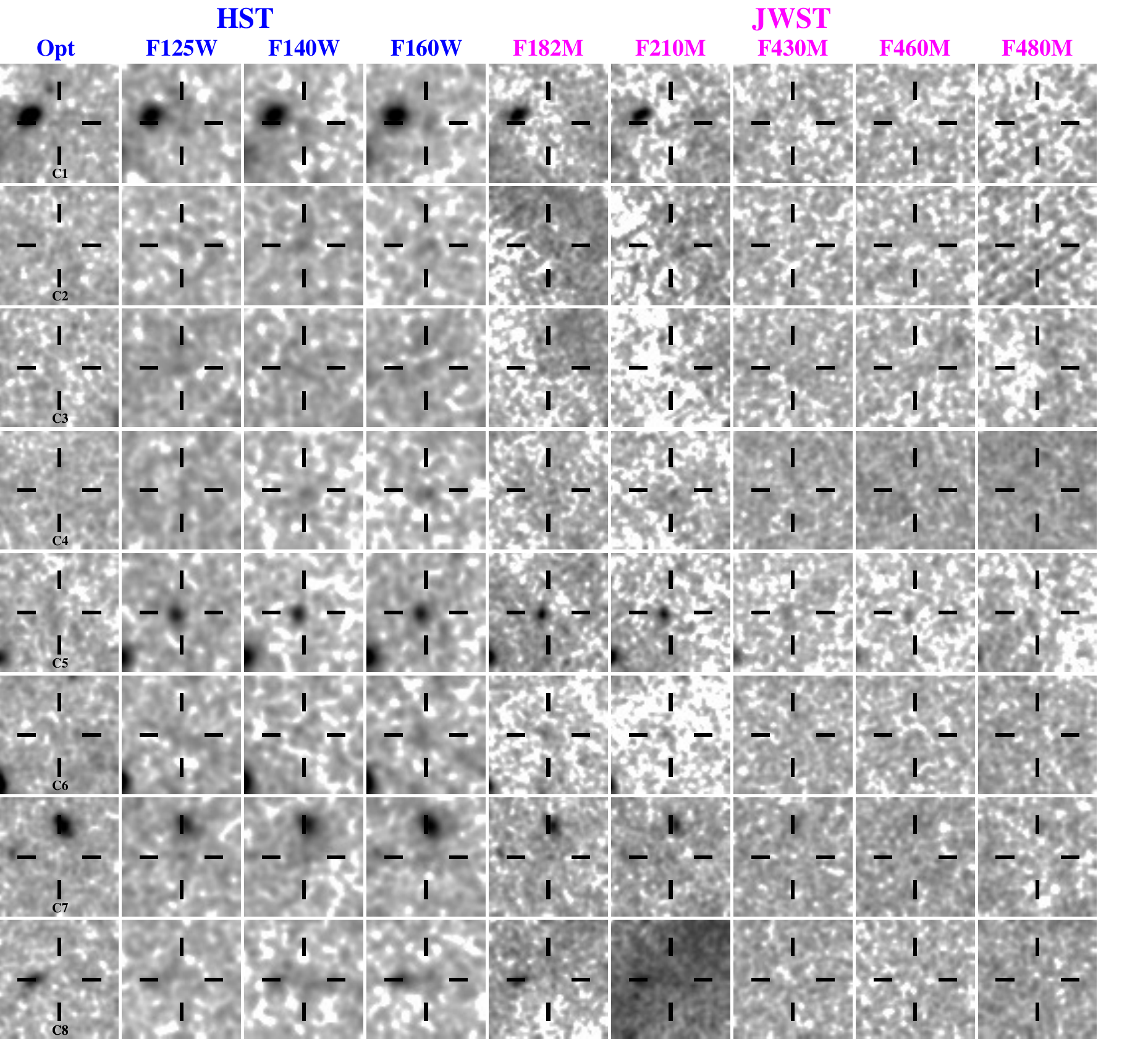}
\caption{Similar to Figure~\ref{fig:stamph} but showing various
  candidate $z=8$-11 galaxies that had been previously reported in the
  literature but which were not recovered using the current selection
  criteria.  While six of the eight candidate show $\geq$2$\sigma$
  detections in the F182M and F210M bands, two candidates notably show
  no apparent flux in the new data from {\it JWST} and may be spurious.  The
  characteristics of the sources we derive from SED fits to our
  photometry are presented in
  Table~\ref{tab:litsample}.\label{fig:stampl}}
\end{figure*}

Postage stamps of the eight candidates are shown in
Figure~\ref{fig:stampl}, and only three of the eight show clear $\geq$
2$\sigma$ detections in both the F182M and F210M images.  While the
NIRCam imaging observations nevertheless support the reality of three
of the remaining five, the NIRCam imaging observations provide no
support for the reality of two of the candidates C1 and C8.  Since one
of these candidates C8 was already suspected to show apparent flux from
a diffraction spike and thus likely to be spurious (see discussion in
\citealt{Oesch2013}), the new observations appear to confirm that this
interpretation was correct.  The absence of a detection in the F182M
and F210M bands for the second candidate C1 is surprising given the
apparent detection of the source at $>$3.5$\sigma$ in the WFC3/IR
F140W and F160W bands \citep{Ellis2013,Oesch2013}.  However, the lack
of apparent flux in the new NIRCam data suggests this source may have either been spurious or variable.

\begin{figure*}
\centering
\includegraphics[width=2\columnwidth]{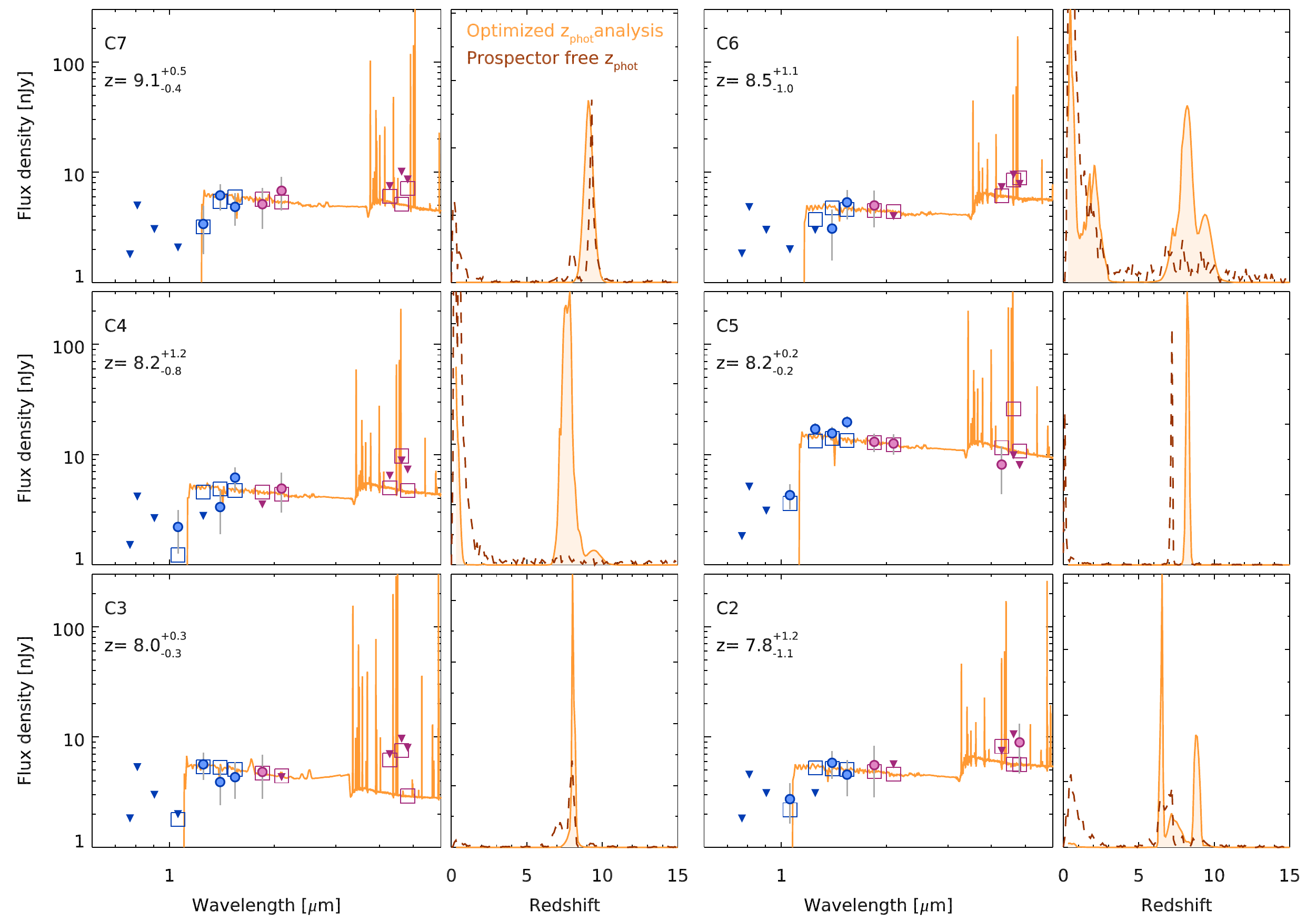}\caption{Similar to Figure~\ref{fig:sedfit3} but for various candidate
  $z=8$-11 galaxies which had previously been reported in the
  literature but are not recovered using the selection criteria used
  here.  The characteristics of the sources we derive from SED fits to
  our photometry are presented in
  Table~\ref{tab:litsample}.\label{fig:sedfitl}}
\end{figure*}

Using the same SED templates from EAzY \citep{Brammer2008} to compute
the redshift likelihood distribution for the six sources that appear
to be real on the basis of the new NIRCam observations, we present our
results in Figure~\ref{fig:sedfitl}.  Our likelihood analysis suggests
that five of the six remaining candidates (C2, C3, C4, C5, C7) are
very likely at $z\geq 8$, but with larger uncertainties regarding the
nature of C6 earlier reported by \citet{Bouwens2011_Nature}.

Of the five candidates that appear likely to be at $z\geq8$, three did
not make it into our selection due to the F182M+F210M medium band
fluxes' showing less than $2\sigma$ detections.  A fourth candidate C5
did not make it into our selection because it was detected at
$>$5$\sigma$ significance in the $Y_{105}$ band and thus excluded.  A
fifth candidate C7 was excluded because it did not make it into the
SExtractor catalog we used to assemble our $z\geq8$ selection over the
HUDF, but appears to otherwise satisfy our selection criteria.

\begin{table*}
\centering
\caption{$z\geq 8.5$ Candidate Galaxies Previously Identified over the HUDF/XDF}
\label{tab:litsample}
\begin{tabular}{c|c|c|c|c|c|c|c|c} \hline
     &  &     &            & $m_{UV}$ & Lyman Break & $\Delta \chi^2$$^{a,d}$ &    &  \\
 ID &  RA & DEC & $z_\mathrm{phot}$$^a$ & [mag]$^b$ & [mag]$^c$ & & p($z>$5.5)$^a$ & Lit$^{1}$ \\\hline\hline
C1 & 03:32:38.95 & $-$27:47:11.7 & --- & $>$30.0 & --- & --- & --- & E13\\
C2 & 03:32:39.44 & $-$27:46:32.0 & 7.8$_{-1.1}^{+1.2}$ & 29.8$\pm$0.5 & $>$1.2 & $-$3.7 & 0.901 & O13\\
C3 & 03:32:38.27 & $-$27:45:56.4 & 8.0$_{-0.3}^{+0.3}$ & 29.8$\pm$0.5 & $>$1.9 & $-$12.0 & 0.994 & B15\\
C4 & 03:32:43.69 & $-$27:46:40.9 & 8.2$_{-0.8}^{+1.2}$ & 29.4$\pm$0.7 & 0.0$\pm$1.5 & $-$0.1 & 0.712 & B11\\
C5 & 03:32:37.79 & $-$27:46:00.4 & 8.2$_{-0.2}^{+0.2}$ & 28.2$\pm$0.1 & 1.5$\pm$0.2 & $-$40.8 & 1.000 & M13,B15\\
C6 & 03:32:35.43 & $-$27:47:33.8 & 8.5$_{-1.0}^{+1.1}$ & 29.6$\pm$0.3 & $>$0.9 & $-$0.2 & 0.546 & B11\\
C7 & 03:32:43.45 & $-$27:46:54.9 & 9.1$_{-0.4}^{+0.5}$ & 29.6$\pm$0.4 & $>$1.3 & $-$12.7 & 0.999 & E13,M13,O13\\
C8 & 03:32:41.07 & $-$27:47:30.7 & --- & $>$30.0 & --- & --- & --- & E13\\
\hline\hline
\end{tabular}
\\\begin{flushleft}
$^1$ B11 = \citet{Bouwens2011_Nature}, E13 = \citet{Ellis2013}, M13 = \citet{McLure2013}, O13 = \citet{Oesch2013}, B15 = \citet{Bouwens2015_LF}\\
$^a$ Derived using EAzY \citep{Brammer2008}\\
$^b$ Derived using the flux in the $H_{160}$, $HK_{182}$, and $K_{210}$ bands for sources in our $z\sim8$-9, $z\sim10$-11, and $z\sim12$-13 samples to probe the $UV$ luminosity at $\approx$1600\AA$\,$rest-frame. \\
$^c$ Amplitude of the nominal Lyman breaks for these $z\geq8$ galaxy candidates.\\
$^d$ $\chi^2_{best,z>5.5} - \chi^2_{best,z<5.5}$\\
\end{flushleft}
\end{table*}

\end{document}